\documentclass[twocolumn]{aastex631}
\usepackage{newtxtext,newtxmath}
\usepackage[T1]{fontenc}

\usepackage{graphicx,graphics}	
\usepackage{comment}
\usepackage{color,epsfig}
\usepackage{amsmath}	
\usepackage{amssymb}	
\usepackage{psfrag}
\usepackage[T1]{fontenc}
\usepackage{ebgaramond}
\usepackage{newtxmath}
\usepackage{booktabs} 
\usepackage{longtable}
\usepackage{supertabular}
\usepackage{float}
\usepackage{multirow}

\let\tablenum\relax
\usepackage{siunitx}

\usepackage{textcomp}

\def\n{\hat{\bm{n}}}

\newcommand{\U}{\boldsymbol{U}}

\newcommand{\bfl}{\boldsymbol{\ell}}

\newcommand{\Hi}{H\,\textsc{i}}

\def\a{{\bm{a}}}

\def\v{\bm{v}}

\def\V{\mathcal{V}}
\def\N{\mathcal{N}}
\def\S{{\mathcal S}}
\def\u{{\bf U}}

\begin{document}

\title{The  Epoch of Reionization 21-cm  Bispectrum at $z=8.2$ from MWA data III: Improved Constraints Using Multiple Pointings}

\correspondingauthor{Md Sirajul Basir}
\email{bmdsirajul@gmail.com}

\author[0009-0001-9269-5414]{Md Sirajul Basir}
\affiliation{Department of Physics, Indian Institute of Technology Kharagpur, Kharagpur 721 302, India}

\author[0000-0003-1629-3357]{Sukhdeep Singh Gill}
\affiliation{Department of Theoretical Physics, Tata Institute of Fundamental Research, Mumbai 400005, India}

\author[0000-0002-2350-3669]{Somnath Bharadwaj}
\affiliation{Department of Physics, Indian Institute of Technology Kharagpur, Kharagpur 721 302, India}

\author[0000-0003-1206-8689]{Khandakar Md Asif Elahi}
\affiliation{Raman Research Institute, C. V. Raman Avenue, Sadashivanagar, Bengaluru 560080, India}
\affiliation{School of Physical Sciences, National Institute of Science Education and Research, Bhubaneswar 752 050, India}

\author[0000-0002-7737-1470]{Shiv K. Sethi}
\affiliation{Raman Research Institute, C. V. Raman Avenue, Sadashivanagar, Bengaluru 560080, India}

\author[0000-0002-6216-2430]{Akash Kumar Patwa}
\affiliation{Raman Research Institute, C. V. Raman Avenue, Sadashivanagar, Bengaluru 560080, India}

\begin{abstract}
We analyze zenith-pointing $(\delta = -26.7^{\circ})$ Murchison Widefield Array (MWA) drift-scan observations covering $349.0^{\circ} \le \alpha \le 70.0^{\circ}$, with $163$ pointing centers (PCs) at  $0.5^{\circ}$  interval. The central frequency  $\nu_c=154.2 \,{\rm MHz}$ corresponds to neutral hydrogen (\Hi) 21-cm radiation from $z=8.2$.   Considering the Epoch of Reionization (EoR) \Hi ~ 21-cm  bispectrum, we quantify the level of foreground contamination as a function of $\alpha$ using $\Delta^3_A$, the mean-cubed angular fluctuation of the sky brightness temperature. We also estimate $\Delta^3$, the mean-cubed 21-cm brightness temperature fluctuations, after applying a smooth component filter (SCF) to mitigate the foregrounds. For both, the PCs around $\alpha \approx 50^{\circ}$ are found to be severely contaminated by the bright extended source Fornax~A. Considering $\Delta^3$, most of the PCs in the range $349^{\circ}\lesssim\alpha\lesssim10^{\circ}$ and a few PCs in the range $65^{\circ}\lesssim\alpha\lesssim70^{\circ}$ are consistent with the $2 \sigma$  expected statistical fluctuations arising from system noise. We use these PCs to constrain $\Delta^3$ of the EoR 21-cm signal to lie within $0 \pm 2 \sigma$. The tightest constraint from a single PC occurs at $\alpha = 69^{\circ}$, with $2\sigma = (8.50 \times 10^3)^3~{\rm mK^3}$, and we obtain an even tighter constraint $2\sigma_c=(4.89\times10^3)^3~{\rm mK^3}$ by combining $99$  PCs, both being at $k_1 = 0.281~{\rm Mpc}^{-1}$ with all triangle shapes collapsed. This is the tightest constraint on the EoR 21-cm bispectrum to date.
\end{abstract}

\keywords{Reionization (1383); Non-Gaussianity (1116); Observational cosmology (1146); \Hi~line emission (690); Diffuse radiation (383); Astronomy data analysis (1858); Radio interferometry (1346); Interferometric correlation (807); Extragalactic radio sources (508) }


\vspace{0.3cm}
\section{Introduction}

The Epoch of Reionization (EoR), during which hydrogen in the inter-galactic medium (IGM) became ionized from a neutral state, marks a very crucial yet unexplored phase in the cosmic evolution. The redshifted 21-cm signal from the hyperfine transition of neutral hydrogen (\Hi) provides a powerful probe of this era, offering insights into the formation and properties of the first stars and galaxies. Radio interferometers, such as the upgraded Giant Metrewave Radio Telescope (uGMRT)\footnote{\url{http://www.gmrt.ncra.tifr.res.in/}} \citep{Swarup_1991,Gupta_2017}, Murchison Widefield Array (MWA)\footnote{\url{https://www.mwatelescope.org/}} \citep{tingay13}, Low Frequency Array (LOFAR)\footnote{\url{https://www.astron.nl/telescopes/lofar/}} \citep{haarlem}, Hydrogen Epoch of Reionization Array (HERA)\footnote{\url{https://reionization.org/}} \citep{DeBoer_2017}, and the upcoming Square Kilometre Array (SKA)-Low\footnote{\url{https://www.skao.int/}} \citep{Koopmans_2015, EoR_SWG_arXiV_2026} are making a significant observational effort to detect the 21-cm signal through measurement of its power spectrum (PS). However, detection has not yet been possible due to the presence of various galactic and extra-galactic foregrounds, which are 4-5 orders of magnitude stronger than the expected signal \citep{ali, Bernardi2009, Ghosh2012}. Currently, we only have a few upper limits on the mean-squared brightness temperature fluctuations, with the best one from the HERA experiment, which reported $\Delta^2 \le (21.4)^2 {\rm mK^2}$ at $k=0.34 ~h~{\rm Mpc^{-1}}$ at $z=7.9$ \citep{HERA2023}. The large data volume expected from the upcoming SKA will significantly improve the constraints and may enable the detection of the signal.

The power spectrum (PS) is a sufficient statistic only for Gaussian random fields. However, the 21-cm signal originating from the EoR is highly non-Gaussian due to the emergence and growth of ionized regions \citep{BP2005,mondal2015, Majumdar_2018}. Consequently, higher-order statistics are required to capture the extra information present in the signal. The bispectrum (BS) is the lowest-order statistic sensitive to such non-Gaussian features. The 21-cm bispectrum has been extensively studied using theoretical models  \citep{Yoshiura_2015, Shimabukuro_2016,Majumdar_2018, Majumdar_2020, trott2019,Hutter_2019, Saxena_2020, Watkinson_2021,Murmu:2023fgv, gill_eormulti, raste_2023}, demonstrating its capability to extract valuable information about the EoR beyond what can be accessed through the power spectrum alone. 

Observational constraints on the EoR 21-cm bispectrum still remain in their preliminary stages. \citet{trott2019} provided a preliminary upper limit of $\sim 10^{12} \text{ mK}^3 \text{Mpc}^6$ at large scales, analyzing 21 hours of $167-197$ MHz $(z = 6.2-7.5)$ MWA Phase~II EoR project data to estimate the 21-cm BS for a few triangle configurations.
A sequential series of works \citep{Gill_2025_mabs, Gill_2024_2d3vc, Gill_2025_mwa1, gill2026epochreionization21cm} has recently led to significant advancements in the observational study of the 21-cm BS. \citet{Gill_2024_2d3vc} developed a fast estimator for the angular bispectrum (ABS) from radio-interferometric visibility data at a single frequency. In a follow-up work, \citet{Gill_2025_mabs} generalized it to multi-frequency observations and developed \texttt{AMBER} estimator, which computes the multi-frequency angular bispectrum (MABS) and the 3D 21-cm bispectrum from visibility data. This estimator was used by \citet{Gill_2025_mwa1} to analyze 17 minutes of a single pointing center (PC) of MWA Phase~II drift-scan observation over the redshift range $z=[7.4, 9.2]$, with a central redshift of $z=8.2$ which corresponds to central frequency $\nu_c=154.2\,{\rm MHz}$. The comprehensive work provided 21-cm BS estimates across all the possible triangle configurations. They identified a foreground wedge in the cylindrical bispectrum, similar to that in the power spectrum, thereby defining an EoR window. However, a periodic pattern of missing frequency channels in the MWA data caused significant foreground leakage, leaving the results dominated by foregrounds despite the use of foreground avoidance methods. The best $2\sigma$ upper limits on the mean-cubed brightness temperature fluctuations were found to be $\Delta^3_{\rm UL} = (1.81 \times 10^3)^3 \text{ mK}^3$ at $k = 0.008 \text{ Mpc}^{-1}$ and $\Delta^3_{\rm UL} = (2.04 \times 10^3)^3 \text{ mK}^3$ at $k = 0.012 \text{ Mpc}^{-1}$ for equilateral and squeezed triangles, respectively. To mitigate the foregrounds, \citet{gill2026epochreionization21cm} employed the Smooth Component Filtering (SCF) technique, which was previously used by \citet{Elahi_missing} in the context of estimating the power spectrum from the same MWA data. Then they used \texttt{AMBER} on the filtered residual visibility data to estimate 21-cm BS for several available near-linear triangle configurations, and presented an improved upper limit as well as a lower limit on the mean-cubed brightness temperature fluctuations $\Delta^3$. Combining the estimates over all triangle shapes, they reported the most stringent constraints at $k = 0.281 \text{ Mpc}^{-1}$, with a lower limit of $\Delta^3_{\rm LL} = -(1.25 \times 10^4)^3 \text{ mK}^3$ and an upper limit of $\Delta^3_{\rm UL} = (1.22 \times 10^4)^3 \text{ mK}^3$, which are consistent with the expected statistical fluctuations around zero arising from system noise. 

Recently, \citet{sarkar_2026} have presented a comprehensive analysis of all 163 PCs of the MWA drift-scan data. They have studied the variation of the angular power spectrum and the 3D power spectrum of the observed brightness temperature fluctuations as a function of right ascension (RA), $\alpha$. The study serves as a guide for future EoR observations, identifying sky regions where the foreground contamination is relatively lower. Additionally, the work demonstrates that by incoherently combining the power spectrum estimates from multiple PCs, the system noise contribution can be effectively suppressed, allowing us to significantly tighten the constraints on the 21-cm power spectrum.  The study reports a $2 \sigma$ upper limit of $\Delta^2_{\rm UL}=(98.67)^2 ~ {\rm mK}^2$ at $k=0.156~{\rm Mpc}^{-1}$, obtained by incoherently combining $23$ PCs.  This is currently the tightest upper limit at $z \approx 8$ from MWA. In a more recent work, \citet{yoshiora2026} have analyzed drift-scan data observed by MWA in the lowest frequency band, spanning 73.6--104.3~MHz, proposing the sky region around (RA, Dec) $= (10^{\circ}\text{-- }40^{\circ}, -26.8^{\circ})$ as a candidate region for future MWA ultra-low observations targeting Cosmic Dawn.

In the present work, we provide a parallel comprehensive analysis of the same dataset used by \citet{sarkar_2026} considering the bispectrum.  Our motivation for this work is threefold. Firstly, our objective is to measure the variation of the foreground contamination in different regions of the sky under consideration. We estimate the angular bispectrum (ABS) for each of the PCs following the methodology developed by \citet{Gill_2024_2d3vc} and show the variation of the mean-cubed angular fluctuations in the observed brightness temperature $(\Delta^3_A)$ as a function of $\alpha$. We expect these findings to guide the selection of optimal target fields for future EoR 21-cm observations. Secondly, we aim to measure the variation of the 3D spherical bispectrum as a function of $\alpha$. Following \citet{gill2026epochreionization21cm}, we apply SCF to mitigate the foregrounds and estimate the mean-cubed brightness temperature fluctuation, $\Delta^3$, for all the PCs separately.  Since the observations are not deep enough to detect the 21-cm signal, the measured $\Delta^3$ is consistent with the statistical fluctuations arising from system noise in some regions and dominated by residual foreground contamination in others. This provides additional guidance to identify the heavily contaminated PCs and select the PCs where the estimated $\Delta^3$ is consistent with system noise as favorable target fields for future SKA-Low observations to detect the EoR 21-cm signal. Finally, we expect the system noise to go down if we combine the estimates from multiple PCs. As a third objective, we incoherently combine the estimates of $\Delta^3$ from different PCs to suppress the system noise contribution and impose tighter constraints on the 21-cm signal.    

The paper is organized as follows. In Section \ref{data}, we describe the MWA drift-scan observational data and the simulation of system noise. Section \ref{sec:methodology} presents the methodology, while the results are reported in Section \ref{sec:results}. Finally, Section \ref{sec:sum} concludes the paper with a summary of our findings. Throughout this paper, we adopt the cosmological parameters from \citet{planck2020}.

\section{Data}
\label{data}

In this paper, we used the data from the MWA Phase~II drift-scan observations (Project ID: G0031), as described in \citet{Patwa_2021}.  The same data were previously used by \citet{chatterjee_2024}, \citet{Elahi_missing} and \citet{sarkar_2026} for estimation of the multi-frequency angular power spectrum (MAPS) and 3D power spectrum, and also by \citet{Gill_2025_mwa1, gill2026epochreionization21cm} to estimate the multi-frequency angular bispectrum (MABS) and 3D bispectrum. The observations have a fixed declination of $\delta=-26.7^{\circ}$, which corresponds to the zenith. The drift scan spans a right ascension (RA) range from $\alpha=349.0^{\circ}$ to $70.0^{\circ}$ (total span of $81^{\circ}$), sampled at an interval of $\Delta\alpha=0.5^{\circ}$, resulting in a total of $163$ pointing centers (PCs). The visibility data were collected for $10$ consecutive nights from October $3$rd to $12$th, $2016$, with an observing time of $5$ hr $24$ minutes per night. The observations were carried out at a central frequency of $\nu_c = 154.2\,\mathrm{MHz}$ with $N_c = 768$ channels of resolution $\Delta \nu_c = 40\,\mathrm{kHz}$, covering a total bandwidth of $B_{\mathrm{bw}} = 30.72\,\mathrm{MHz}$. This corresponds to probing the \Hi\, 21-cm line over the redshift range $7.4 \leq z \leq 9.2$, with a central redshift of $z = 8.2$.  The bandwidth is divided into 24 coarse bands of $1.28\,\mathrm{MHz}$ (32 channels each). Due to the design of the correlator in the MWA Phase~II \citep{Prabu2015}, the four edge channels and the central channel in each coarse band are flagged. Additional channels may be flagged at each baseline to mitigate radio frequency interference (RFI). After flagging and calibrating the data from each night separately, the data are averaged over all nights. Owing to the loss of the first 2 hr of data on the sixth night, the number of nights is $N_{\mathrm{nights}} = 10$ for PCs with $\alpha > 18.5^\circ$, and $N_{\mathrm{nights}} = 9$ otherwise. The final measurement set (MS) for each PC contains visibilities at 11 time stamps, each with an effective integration time of $t_{\mathrm{int}} = N_{\mathrm{nights}} \times 10\,\mathrm{s}$. Most of the baselines in the data ($\sim 99\%$) lie within $U = 250$ \citep{chatterjee_2024}. We, therefore, restrict our analysis to the baselines in the range $5 \le U \le 270$. For each baseline $\u$ and frequency $\nu$, visibilities are measured in two orthogonal polarizations, XX and YY, denoted as $\mathcal{V}_{\mathrm{XX}}(\u,\nu)$ and $\mathcal{V}_{\mathrm{YY}}(\u,\nu)$. These are treated as independent measurements and combined during the gridding process.

To estimate the system noise contribution to the bispectrum (BS), we perform dedicated noise-only simulations. The real and imaginary parts of the visibilities are modeled as independent Gaussian random variables with zero mean and standard deviation
\begin{equation}
\sigma_N = \frac{1}{\eta_Q} \frac{\sqrt2 k_B T_{\mathrm{sys}}}{A_{\mathrm{eff}} \sqrt{\Delta \nu_c \, t_{\mathrm{int}}}} \, ,
\end{equation}
\citep{Thompson_1986, chatterjee_2024}. Here, $A_{\mathrm{eff}} = 21.4\,\mathrm{m}^2$, $T_{\mathrm{sys}} = 400\,\mathrm{K}$ (dominated by sky temperature at low frequencies), $\eta_Q = 1$, and $t_{\mathrm{int}} = N_{\rm nights}\times10\,\mathrm{s}$, giving $\sigma_{\rm N}~\approx 60~\mathrm{Jy}/\sqrt{N_{\rm nights}}$ \citep{Patwa_2021}. 

Noise is generated independently for each baseline, frequency channel, timestamp, and polarization, using the same observational setup and flagging as the data. We produce $50$ independent realizations for each PC to estimate the noise-induced uncertainty in the bispectrum measurements.

\section{Methodology}
\label{sec:methodology}
In our analysis, we approximate the MWA primary beam pattern, $\mathcal{A}(\theta)$, as a Gaussian beam,
$\mathcal{A}(\boldsymbol{\theta})=\exp\left(-\theta^2/\theta_0^2\right),$
where $\theta_0 = 0.6\,\theta_F$, with $\theta_F = 24.68^\circ$ being the full width at half maximum (FWHM) of the MWA primary beam \citep{Gill_2024_2d3vc}.

\subsection{Angular Bispectrum}

To begin with a two-dimensional (2D) analysis, we require visibilities $\mathcal{V}(\u_i)$ which quantify the brightness temperature fluctuations on the sky at a fixed frequency.  We introduce a square grid in the baseline $(uv)$ plane with grid spacing $\Delta U_g = \sqrt{\ln 2}/{\pi \theta_0} \approx 1$, and assign each of the visibilities $\mathcal{V}(\mathbf{U}_i)$ to the nearest grid point $\mathbf{U}_g$ using

\begin{equation}
\V_{g}(\nu) = \sum^{N_g(\nu)}_{i}\tilde{w}(\u_g-\u_i) \, \V(\u_i,\nu) \, F(\u_i,\nu).
\label{eq:vg}
\end{equation}
Here $\tilde{w}(\mathbf{U_g} - \mathbf{U}_i) = 1$ if $\u_g$ is the point in the grid closest to $\u_i$, and $0$ otherwise. The factor $F(\u_i, \nu)$ handles flagged data, taking the value 0 if the data at baseline $\u_i$ are flagged, and $1$ otherwise. $N_g$ counts the number of visibilities contributing to each grid point, defined as $N_g(\nu) = \sum_i\tilde{w}(\mathbf{U_g} - \mathbf{U}_i)\,F(\mathbf{U}_i,\nu)$. We collapse four unflagged frequency channels near the central frequency channel, spanning 153.68--153.78\,MHz, into a single frequency channel and grid the data according to Equation \ref{eq:vg} keeping $\nu$ constant.

Then we use the visibility-based angular bispectrum estimator \citep{Gill_2024_2d3vc}, implemented using FFT-based techniques, which computes the angular bispectrum (ABS), $B_A$, through the three-visibility correlation
\begin{equation}
\begin{aligned}
&& \langle \V(\u_1) \V(\u_2) \V(\u_3+\Delta \u)\rangle =
 (\pi\theta_0^2 Q^3/3)  \\
&&\times
  \exp\left [-{\pi^2\theta_0^2\Delta U^2/3} \right ] 
 B_A({\ell_1}, {\ell_2}, {\ell_3}) \,.
\label{eq:abs}
\end{aligned}
\end{equation}
Here, $\u_1+\u_2+\u_3=0$, i.e., the baselines form a closed triangle, $\Delta\u$ denotes the deviation from a closed triangle configuration, and $\bfl = 2\pi\u$. The Rayleigh-Jeans factor $Q=2k_B/\lambda^2$ converts brightness temperature to specific intensity. We re-parameterize $B_A(\ell_1,\ell_2,\ell_3)$ as $B_A(\ell_1,\mu_A, t_A)$, where 
$\ell_1\ge\ell_2\ge\ell_3$ is assumed, with $\ell_1$ quantifying the size of the triangle, while the parameters $t_A=\ell_2/\ell_1$ and $\mu_A=-(\boldsymbol{\ell_1}\cdot\boldsymbol{\ell_2})/(\ell_1\ell_2)$ together specify the triangle shape. The readers are referred to \citet{bharadwaj_2020} for a detailed description of the parameterization adopted here.

\subsection{MABS and 3D Bispectrum}

To estimate the 3D bispectrum, we grid the data in a similar manner as the ABS, with an additional axis for frequency, as presented in Equation \ref{eq:vg}. 
Following \citet{gill2026epochreionization21cm} and \citet{Elahi_missing},  we apply Smooth Component Filtering (SCF)  to mitigate the foreground contamination. This  convolves the gridded visibility data  with a Hann window function $H(n)$ to get its smooth component 
\begin{equation}
    \V^s_{g}(\nu_n) = (  \V_{g} \ast H )(\nu_n) = \sum_{m=1}^{N_c} \V_{g}(\nu_{m}) H(n-m)\,
    \label{eq:conv}
\end{equation}
Here, $N$ determines the half-width of the smoothing kernel, while the full width of the window is $2N+1$. We use $N=50$, corresponding to a smoothing half-width of
$N\Delta\nu_c = 2\,\mathrm{MHz}$, for reasons discussed in \citet{Elahi_missing}. We then subtract the smoothed component to get the residual filtered component,
\begin{equation}
    \V^f_{g}(\nu) =\V_{g}(\nu) - \V^s_{g}(\nu) \,.
    \label{eq:res}
\end{equation}
To incorporate the correct normalization in the presence of flagged channels, we divide $\V^f_{g}(\nu)$ by $(F \ast H)(\nu)$, where, as mentioned earlier, $F(\nu)$ encodes the flagging information. As the convolution fails near the band edges, after the convolution, we discard $N=50$ channels from both of the edges. The resulting bandwidth becomes $B_{\rm bw}^f = B_{\rm bw} - 2N\Delta\nu_c= 26.72\,{\rm MHz}$. The readers are referred to \citet[Section 6]{Elahi_missing} for a detailed description of the technique.

We next use \texttt{AMBER} \citep{Gill_2025_mabs} to compute the MABS, $B(\ell_1, \ell_2, \ell_3, \nu_1, \nu_2,\nu_3)$. This evaluates three-visibility correlations 

\begin{equation}
\begin{aligned}
&& \langle \V(\u_1,\nu_1) \V(\u_2,\nu_2) \V(\u_3+\Delta \u,\nu_3)\rangle =
 (\pi\theta_0^2 Q^3/3)  \\
&&\times
  \exp\left [-{\pi^2\theta_0^2\Delta U^2/3} \right ] 
 B_A({\ell_1}, {\ell_2}, {\ell_3},\nu_1,\nu_2,\nu_3) \,.
\label{eq:mabs}
\end{aligned}
\end{equation}
Here, all the symbols carry the same meaning as in the ABS formalism (eq.~\ref{eq:abs}). $Q$ and $\theta_0$ are evaluated at the central frequency $\nu_c$. Furthermore, assuming ergodicity along the line-of-sight (LoS) direction, the MABS can be expressed as a function of the frequency separations, $B_A(\ell_1, \ell_2, \ell_3, \Delta\nu_1, \Delta\nu_2)$, where $\Delta\nu_1=\nu_1-\nu_3$ and $\Delta\nu_2=\nu_2-\nu_3$.

We obtain the 3D cylindrical bispectrum by performing a 2D Fourier transform along $(\Delta\nu_1, \Delta\nu_2)$, following \citet{BP2005},
\begin{equation}
    \begin{aligned}
 &B(k_{1 \perp},k_{2 \perp},k_{3 \perp},k_{1\parallel},k_{2\parallel}) = 
 r^4 r'^2 \int d(\Delta \nu_1) \int d(\Delta \nu_2)  
\\&\times \exp{[-ir'(k_{1 \parallel}\Delta \nu_1+k_{2 \parallel}\Delta \nu_2)]} \, B_A(\ell_1, \ell_2, \ell_3, \Delta \nu_1, \Delta \nu_2)~.
\end{aligned}
\label{eq:3dbs_mabs}
\end{equation}
where $r$ is the comoving distance corresponding to the central frequency $\nu_c$, calculated from the redshift $z_c = \frac{1420\,\mathrm{MHz}}{\nu_c} - 1$, and $r' = dr/d\nu$. For the present analysis at $z_c = 8.2$, we have $r \simeq 9210\,\mathrm{Mpc}$ and $r' \simeq 17\,\mathrm{Mpc\,MHz^{-1}}$.

The cylindrical bispectrum $B(k_{1\perp}, k_{2\perp}, k_{3\perp}, k_{1\parallel}, k_{2\parallel})$ depends on five independent variables, where $\mathbf{k}_\perp = \boldsymbol{\ell}/r$ and $k_\parallel$ denote the components of $\mathbf{k}$, perpendicular and parallel to the LoS direction, respectively. The parallel components also follow the closed triangle condition, $k_{1\parallel}+k_{2\parallel}+k_{3\parallel}=0$. The Fourier transform in Equation \ref{eq:3dbs_mabs} introduces ripple-like oscillations in the $(k_{1\parallel}, k_{2\parallel})$ plane due to the sharp truncation of the MABS at the boundaries of $(\Delta\nu_1, \Delta\nu_2)$. To mitigate these edge effects, we apply a 2D Blackman--Nuttall window function \citep{Nuttall_1981}, which smoothly tapers the MABS near the boundaries and suppresses the resulting ripples.

Finally, we spherically bin the 3D cylindrical bispectrum, $B(k_{1\perp}, k_{2\perp}, k_{3\perp}, k_{1\parallel}, k_{2\parallel})$, to obtain the spherical bispectrum, $\bar{B}(k_1, k_2, k_3)$, where $k_1 = \sqrt{k_{1\perp}^2 + k_{1\parallel}^2}$, and similarly for $k_2$ and $k_3$. SCF is expected to suppress the 21-cm signal below a characteristic scale $k_{\parallel} = [k_{\parallel}]_{\rm F}$, which is related to the smoothing scale. \citet{sarkar_2026} has shown through simulations that SCF does not result in a significant signal loss above $k_{\parallel}=0.135~{\rm Mpc}^{-1}$. Further, SCF becomes less effective in mitigating foreground contamination at large $k_{\perp}$. Therefore, as in \citet{gill2026epochreionization21cm}, we restrict our analysis to the region defined by $k_{\parallel} \geq 0.135~{\rm Mpc}^{-1}$ and $k_{\perp} \leq 0.026~{\rm Mpc}^{-1}$, and evaluate the spherical BS using only the modes within this $(k_{\perp}, k_{\parallel})$ domain. In addition, we re-parameterize $\bar{B}(k_1,k_2,k_3)$ as $B(k_1,\mu,t)$, similar to the ABS parameterization, where $k_1 \geq k_2 \geq k_3$ is assumed. Here, $k_1$ quantifies the size of the triangle, while the parameters $t = k_2/k_1$ and $\mu = (-\mathbf{k_1}\cdot\mathbf{k_2})/(k_1k_2)$ together specify the triangle shape \citep{bharadwaj_2020}.

\section{Results}
\label{sec:results}
\subsection{Angular Bispectrum}

\begin{figure*}[htbp]
  \centering
  \includegraphics[width=0.945\textwidth]{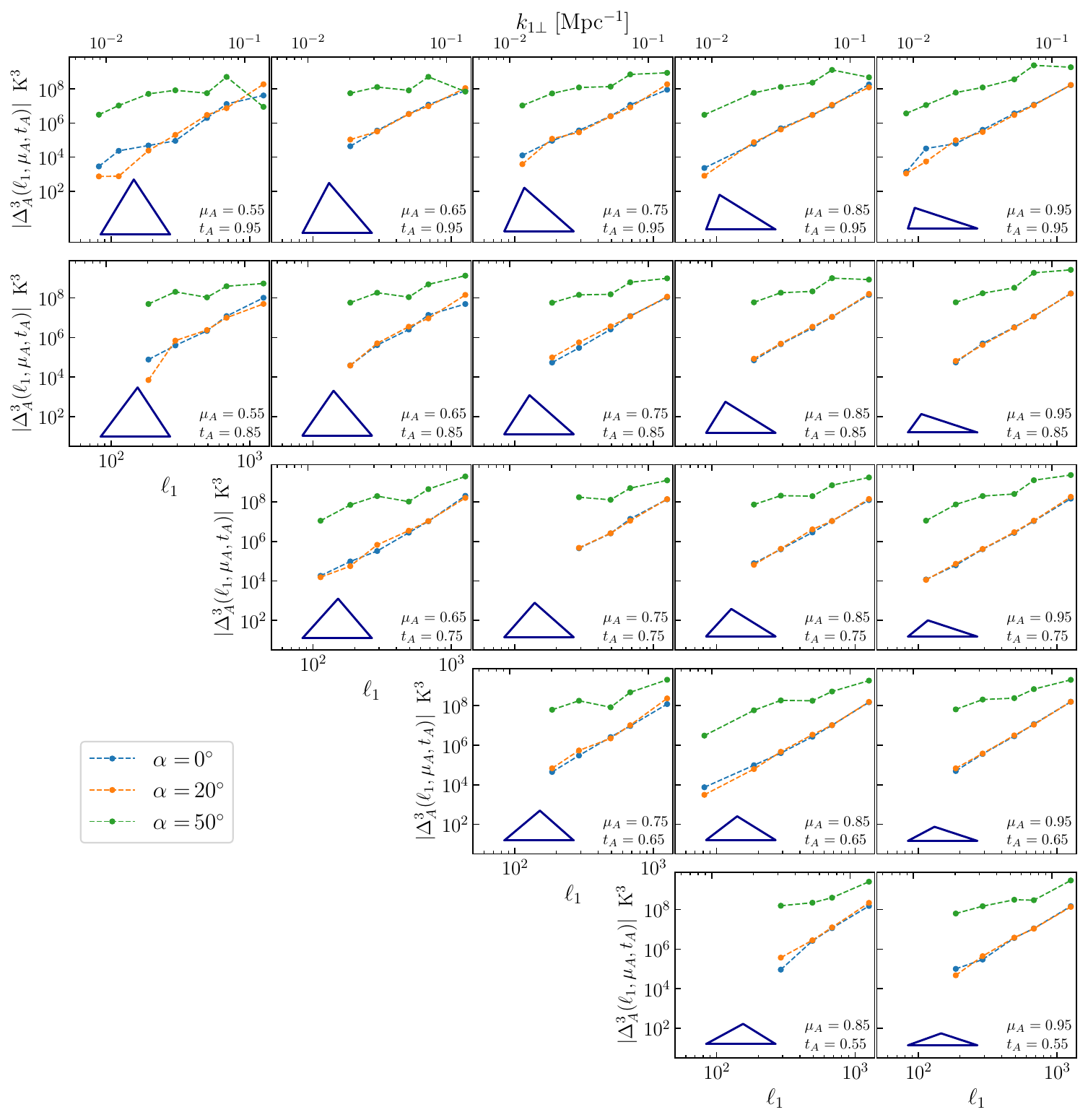}
  \caption{Each panel corresponds to a different set of $(\mu_A, t_A)$ values, spanning the entire allowed range and covering triangles of all possible shapes. In each panel, $|\Delta_A^3(\ell_1, \mu_A, t_A)|$ is shown as a function of $\ell_1$ for three pointing centers, $\alpha = 0^{\circ}, 20^{\circ},$ and $50^{\circ}$, with different colors as indicated in the figure. In the bottom left corner of each panel, we have shown the shape of the corresponding triangle for a fixed value of $\ell_1$. The top axis shows the corresponding value of $k_{1\perp} = \ell_1 / r$, where $r= 9210\,\mathrm{Mpc}$ is the comoving distance to the observed redshift.}
  \label{fig:ABS_all_shape}
\end{figure*}

\begin{figure*}[htbp]
  \centering
  \includegraphics[width=1\textwidth]{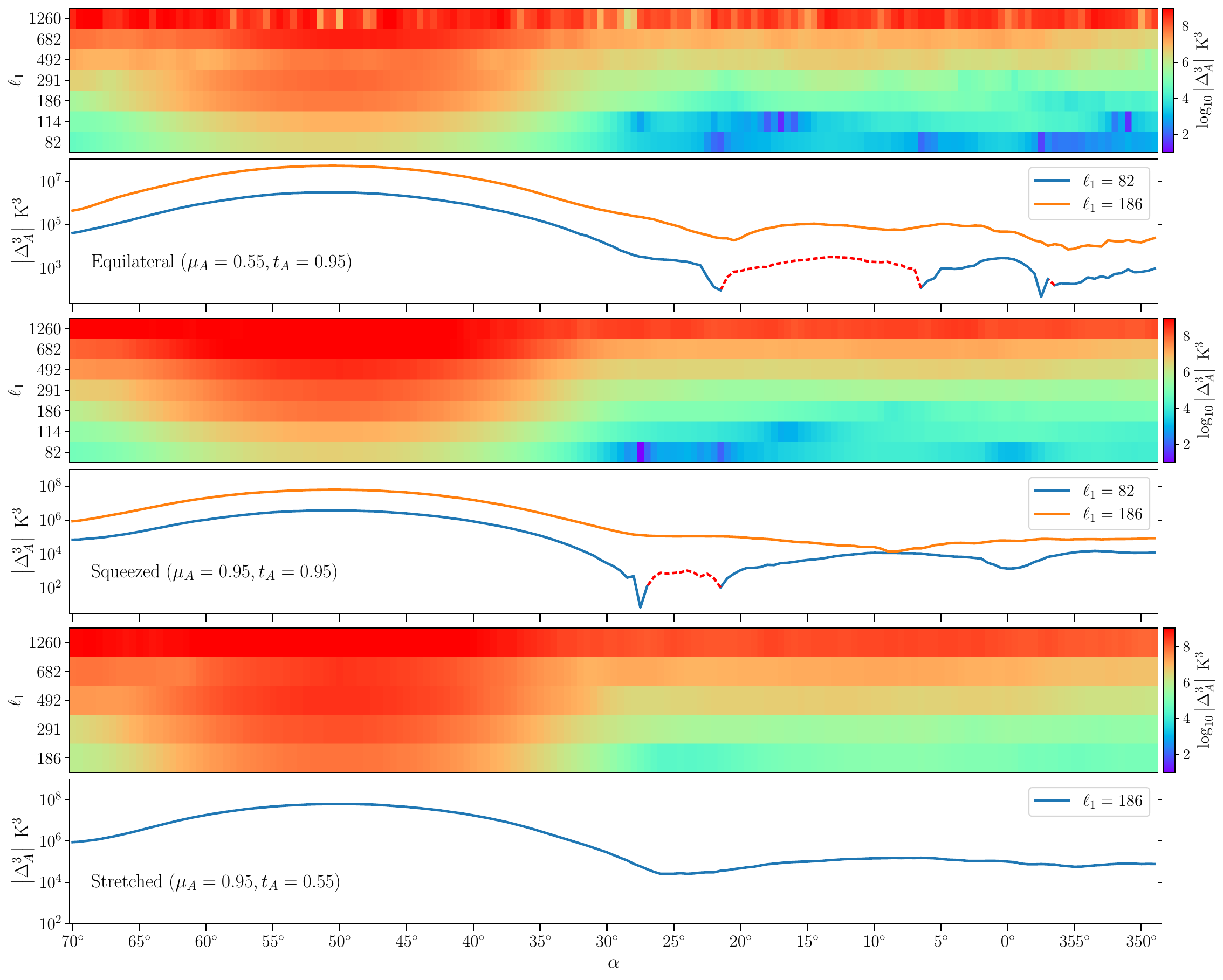} 
   \caption{The variation of the mean-cubed angular fluctuation of the sky brightness temperature, $|\Delta^3_A|$, as a function of the pointing center $\alpha$ for three representative triangle configurations: equilateral (top), squeezed (middle), and stretched (bottom). Each configuration is shown as a 2D heatmap of $\log_{10}|\Delta_A^3|$ as a function of $\ell_1$ and $\alpha$, together with a line plot of $|\Delta_A^3|$ at selected values of $\ell_1$. The red dashed lines in the line plots indicate negative values of $\Delta_A^3$.}
  \label{fig:ABS_heatmap}
\end{figure*}

In this section, we consider  $\Delta^3_A = B_A (\ell_1^2/2\pi)^2$, which quantifies the mean-cubed angular brightness temperature fluctuation of the sky signal, as a function of $\ell_1$. 
Following \citet{Gill_2024_2d3vc}, we have divided the available range $20\le\ell_1\le1570$ into $9$ logarithmic bins of equal spacing, and discarded the two smallest bins $(\ell_1 <80)$ which do not provide reliable results.  Each panel of  Figure~\ref{fig:ABS_all_shape} shows the results for a different triangle shape ($(\mu_A,t_A)$ value), as illustrated (and mentioned) in the figure. Here, the top left panel corresponds to equilateral triangles $((\mu_A,t_A) \approx (0.5,1))$, the top right panel corresponds to squeezed triangles $((\mu_A,t_A) \approx (1,1))$ and the bottom right panel corresponds to stretched triangles $((\mu_A,t_A) \approx (1,0.5))$. The panels together cover triangles of all possible shapes. Each panel  shows $|\Delta^3_A|$ as a function of $\ell_1$   for three different PCs centered at $\alpha = 0^{\circ},\, 20^{\circ}\,{\rm and}\, 50^{\circ}$ respectively. The PC centered at $\alpha=0^{\circ}$ corresponds to  EoR0, which is a well-known field for MWA  EoR observations \citep{Carroll2016, Trott2020, Nunhokee2025}. 

We see that the results for $\alpha=0^{\circ}$ are quite similar to those for  $20^{\circ}$,  which are both quite different from those for $\alpha = 50^{\circ}$. 
At $\alpha = 0^{\circ}\, {\rm and}\, 20^{\circ}$, $|\Delta^3_A|$ increases approximately as $|\Delta^3_A| \propto \ell^4$, although a quantitative power law fit does not work. This scaling indicates that the measured signal is dominated by the Poisson fluctuations of the extragalactic point sources, similar to that seen for the angular power spectrum \citep{Chatterjee_2025,sarkar_2026}. 
However, this trend changes at $\alpha = 50^{\circ}$, where the slope of $|\Delta^3_A|$ flattens, and the  power-law-like behavior is missing. Furthermore, at low $\ell_1$, the magnitude of $\Delta^3_A$  is nearly three orders of magnitude larger compared to $\alpha = 0^{\circ}\, {\rm and}\, 20^{\circ}$. 
For example, considering  squeezed triangles,  at $\ell_1 = 82$  we have  $\Delta^3_A = 1.4 \times 10^3 \, {\rm and} \, 1.1 \times 10^3 \, \mathrm{K}^3$ for $\alpha = 0^{\circ} \, {\rm and} \,  20^{\circ}$, respectively, whereas $\Delta^3_A = 3.7 \times 10^6 \, \mathrm{K}^3$ for $\alpha = 50^{\circ}$. 
We attribute this to the exceptionally bright radio source Fornax~A, located at $(\alpha,\delta)=(50.5^{\circ}, -37.2^{\circ})$, which lies within the main lobe of the MWA primary beam for the PC at $\alpha=50^{\circ}$. Fornax~A is an extended radio source consisting of a central core and two radio lobes, with a total flux density of 750 Jy at 154 MHz \citep{McKinley2025}. The difference in the value of $\Delta^3_A$ across different PCs becomes less prominent as $\ell_1$ increases.

 We now study the dependence of $\Delta^3_A$ on the triangle shapes. First, we note that the available $\ell_1$ range varies with the triangle shape. We obtain estimates over the entire range $80\le\ell_1\le1570$  only for the equilateral and squeezed triangles.  Estimates are not available at low $\ell_1$  for the other shapes, as these do not occur for any combination of the annular rings considered here. We have the smallest range 
 $291\lesssim\ell_1\lesssim1260$, for right-angled triangles  $(\mu_A,t_A)=(0.75,0.75)$ and also for    $(\mu_A,t_A)=(0.85,0.55)$.  To discuss the shape dependence, we focus on $\ell_1\approx492$, which is available for all shapes. For the PC at $\alpha=0^{\circ}$, $\Delta^3_A$ is minimum $(2.0 \times10^6~{\rm K^3})$ for equilateral triangles,  and maximum  $(3.7 \times10^6~{\rm K^3})$ for the squeezed and stretched triangles. 
 For the PC at $\alpha=20^{\circ}$, $\Delta^3_A$ is  minimum   $( 2.3\times10^5~{\rm K^3})$  for the isosceles triangle with $(\mu_A,t_A)=(0.75,0.65)$, and it is    maximum  $(4.1\times10^5~{\rm K^3})$ for $(\mu_A,t_A)=(0.85,0.75)$. For both these PCs, the value of 
   $\Delta^3_A$ exhibits very moderate variation across the different triangle shapes. 
   At $\alpha=50^{\circ}$, $\Delta^3_A$ is  minimum $(5.7\times10^7~{\rm K^3})$ for  equilateral triangle, and maximum $(3.8 \times10^8~{\rm K^3})$ for squeezed triangles. For this PC,  the magnitude of $\Delta_A^3$ decreases rapidly with decreasing $\mu_A$, whereas it exhibits moderate variations with $t_A$.

Figure~\ref{fig:ABS_heatmap} shows the variation of $\Delta^3_A$ as a function of $\alpha$ for three representative triangle shapes: equilateral, stretched, and squeezed. The figure is structured into three distinct pairs of panels, each dedicated to a specific 
shape. For each shape,  the upper panel presents a heatmap illustrating the magnitude of $|\Delta^3_A|$ as a function of both the  $\ell_1$ (y-axis) and the pointing center $\alpha$ (x-axis). This panel provides a visual representation of how the non-Gaussian foreground contamination evolves across varying angular scales and directions on the sky. 
For each shape,  the lower panel displays $|\Delta^3_A|$ as a function of $\alpha$ at a few fixed values of $\ell_1$. These line plots provide a more detailed view of the $\alpha$ dependence, enabling us to identify the region least affected by foregrounds and hence best suited for EoR observations. The severely contaminated PCs can thereby be excluded from future analysis.

Considering the heatmap for equilateral triangles, we see that $|\Delta^3_A|$ increases with $\ell_1$ across all the PCs, consistent with our observations in Figure~\ref{fig:ABS_all_shape}. At low $\ell_1$, $|\Delta^3_A|$ exhibits a noticeable  dependence on $\alpha$, with a prominent peak  around  $\alpha\approx50^{\circ}$, near the location of Fornax~A. As $\ell_1$ increases, this $\alpha$ dependence progressively weakens, and the signal becomes nearly isotropic at the highest $\ell_1$, with only weak, rapidly varying fluctuations remaining throughout the full range of $\alpha$. We can infer that at the higher $\ell_1$, the signal becomes progressively dominated by Poisson fluctuations from extragalactic point sources. The accompanying panel below the heatmap shows two line plots corresponding to $\ell_1 = 82$, the smallest multipole, and $\ell_1 = 186$. For $\ell_1 = 82$, a broad dominant peak is observed around $\alpha = 50^{\circ}$, along with two smaller peaks near $\alpha \approx 14^{\circ}$ and $1^{\circ}$.  Note that  $\Delta^3_A$  has negative values in the range $\alpha=7^{\circ}$ to $21.5^{\circ}$, whereas it is positive elsewhere.  For $\ell_1 = 186$, $\Delta^3_A$ is positive throughout, and it is nearly an order of magnitude larger than that for $\ell_1 = 82$ across the entire $\alpha$ range. The dominant peak around $\alpha\approx50^{\circ}$ remains clearly visible; however, the smaller peaks become much less prominent. The overall variation with $\alpha$  becomes weaker  compared to 
$\ell_1 = 82$.

\begin{figure*}[htbp]
    \centering
    \includegraphics[width=\textwidth]{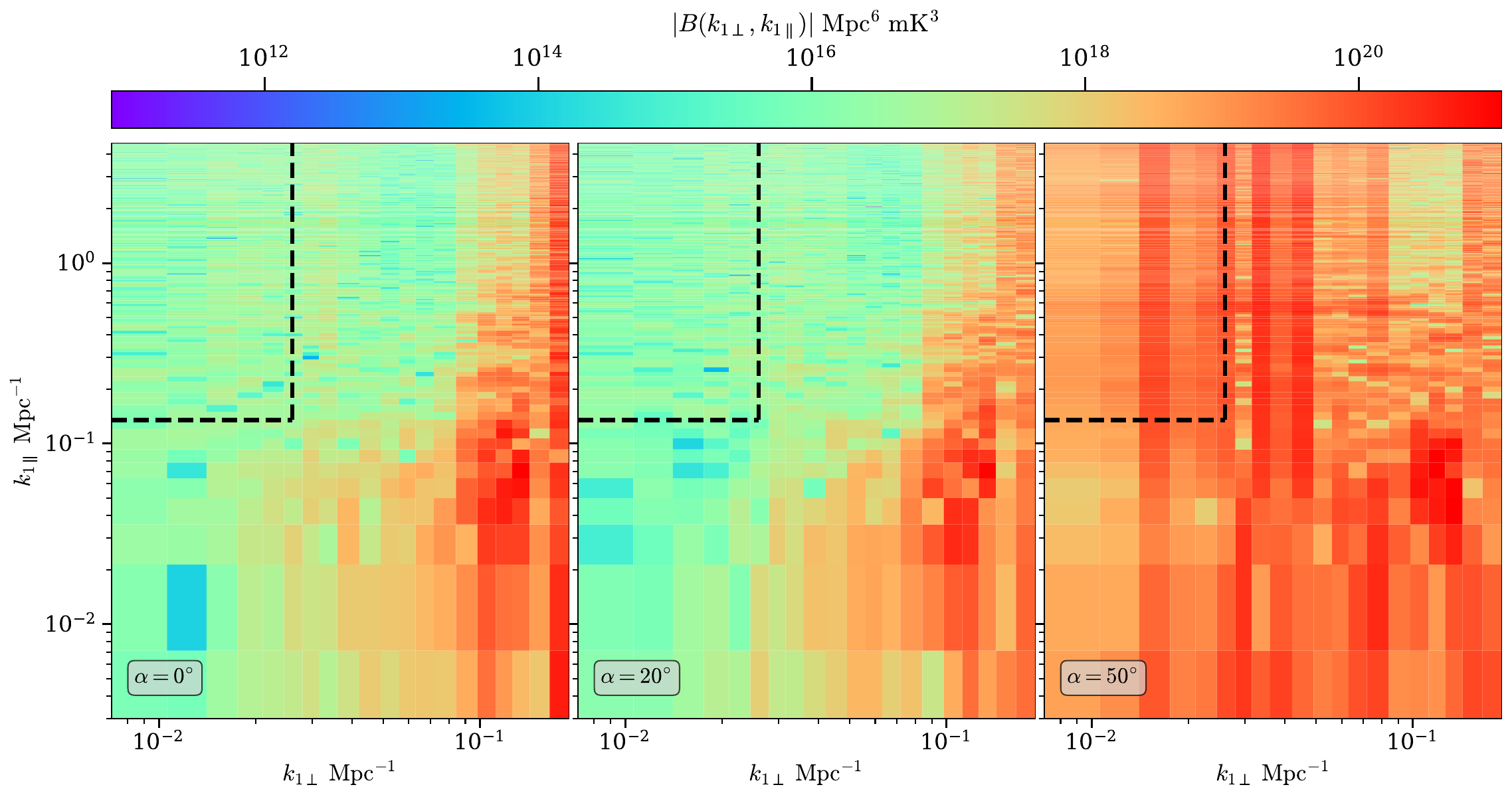}
    \caption{The magnitude of the 3D cylindrical BS $|B(k_{1\perp}, k_{2\perp}, k_{3\perp}, k_{1\parallel}, k_{2\parallel})|$ in units of $\text{Mpc}^{6}\,\text{mK}^{3}$, with $k_{1\parallel} = k_{2\parallel}$, shown as a function of $(k_{1\perp}, k_{1\parallel})$ after applying SCF. An equilateral triangle configuration with $k_{1\perp} \approx k_{2\perp} \approx k_{3\perp}$ is considered here. The three panels represent three representative PCs: $\alpha=0^{\circ}$, $20^{\circ}$, and $50^{\circ}$. The black dashed boxes enclose the region $k_{1\perp} \leq 0.026 \text{ Mpc}^{-1}$ and $k_{1\parallel} \ge 0.135 \text{ Mpc}^{-1}$, highlighting the modes used to evaluate the spherical BS and constrain the EoR 21-cm signal.}
    \label{fig:cylbs_eq}
\end{figure*}

The heatmap corresponding to the squeezed triangle shows overall similar behavior as the equilateral case, with only a few minor differences. As before, a dominant peak is observed around $\alpha \approx 50^{\circ}$. However, at higher values of $\ell_1$, $|\Delta^3_A|$ varies more smoothly with $\alpha$. At the highest $\ell_1$, the fluctuations in $|\Delta_A^3|$ are comparatively weaker than the equilateral configuration and appear to be more uniform across the full range of $\alpha$. The corresponding line plots show behavior broadly consistent with the equilateral case, with a persistent dominant peak near $\alpha \approx 50^{\circ}$. However, the secondary peaks at $\ell_1=82$, are less pronounced. $\Delta^3_A$ takes negative values in the range from $\alpha=22^{\circ}$ to $\alpha=27^{\circ}$, whereas it is positive elsewhere. At $\ell_1=186$, fluctuations with $\alpha$ become substantially smoother, resulting in a more uniform appearance.

For the stretched triangles, we obtain values of $|\Delta^3_A|$ only for $\ell_1 \ge 186$. The corresponding heatmap is very similar to that of the squeezed triangles. We show the line plot at $\ell_1=186$ below. The dominant peak centered around $\alpha \approx 50^{\circ}$ remains clearly visible, while beyond $\alpha \approx 26^{\circ}$, $\Delta^3_A$ becomes nearly flat.

We use noise-only simulations, as described in Section~\ref{sec:methodology}, to estimate the error bars in $\Delta_A^3$ due to system noise. We compute ABS for each noise realization and bin them in the same manner as $B_A(\ell_1,\mu,t)$. The system-noise contribution is estimated from the standard deviation of 50 realizations. The resulting error bars are too small to be visible in both Figure~\ref{fig:ABS_all_shape} and Figure~\ref{fig:ABS_heatmap}. 

Overall, at low $\ell_1$, the region $349^{\circ}\lesssim\alpha\lesssim30^{\circ}$, appears to be comparatively less affected by foreground contamination than the other regions. This indicates that these pointing centers might be more suitable for constraining the 21-cm bispectrum. The findings presented in this section are in qualitative agreement with those reported by \citet{sarkar_2026}, who have considered the $\alpha$ dependence of the angular power spectrum (APS).

\subsection{Cylindrical Bispectrum}

In this section, we present the results for the 3D cylindrical bispectrum. Figure \ref{fig:cylbs_eq} shows the absolute value of the 3D cylindrical bispectrum, $|B(k_{1\perp}, k_{2\perp}, k_{3\perp}, k_{1\parallel}, k_{2\parallel})|$ after applying SCF, for equilateral triangles  ($k_{1\perp} \approx k_{2\perp} \approx k_{3\perp}$) considering  the same three representative PCs as in Figure \ref{fig:ABS_all_shape}. We consider triangle  configurations with $k_{1\parallel} = k_{2\parallel} = -k_{3\parallel}/2$, which  allows us  to show the 3D cylindrical bispectrum using  $\mid B(k_{1\perp},k_{1\parallel}) \mid $, i.e.  a function of $k_{1\perp}$ and $k_{1\parallel}$ alone. These plots serve to visually demonstrate the level of residual foreground contamination after SCF. The  black dashed rectangle in each panel of  
Figure \ref{fig:cylbs_eq} demarcates the region  $k_{\parallel} \geq 0.135~{\rm Mpc}^{-1}$ 
and $k_{\perp} \leq 0.026~{\rm Mpc}^{-1}$.  As mentioned in Section \ref{sec:methodology} and discussed in  \citet{gill2026epochreionization21cm}, we have only used the triangles where all three sides $(\mathbf{k_1},\mathbf{k_2},\mathbf{k_3})$  lie within this region to estimate the spherical bispectrum. Considering the left and middle panels, which correspond to $\alpha=0^{\circ}$  and $20^{\circ}$ respectively, we see that  $\mid B(k_{1\perp},k_{1\parallel}) \mid $ appears to be relatively clean and free of foreground contamination within the chosen rectangular region, whereas the region 
 $k_{\perp} > 0.026~{\rm Mpc}^{-1}$ appears to be contaminated even after SCF. As discussed in 
\citet{gill2026epochreionization21cm}, SCF is not as effective at large $k_{\perp}$ where the foregrounds exhibit rapid frequency variations that are not filtered by the  Hann window. 
In contrast, the right panel $(\alpha = 50^\circ)$ shows substantial foreground contamination both within and outside the rectangular region. We see that for this PC, whose $\alpha$ coincides with that of Fornax~A,  there is a large foreground contamination even after SCF. Since similar behavior is observed for other PCs near $\alpha = 50^\circ$, they must be excluded when estimating the EoR 21-cm BS. 

\begin{figure*}[htbp]
    \centering
    \includegraphics[width=\textwidth]{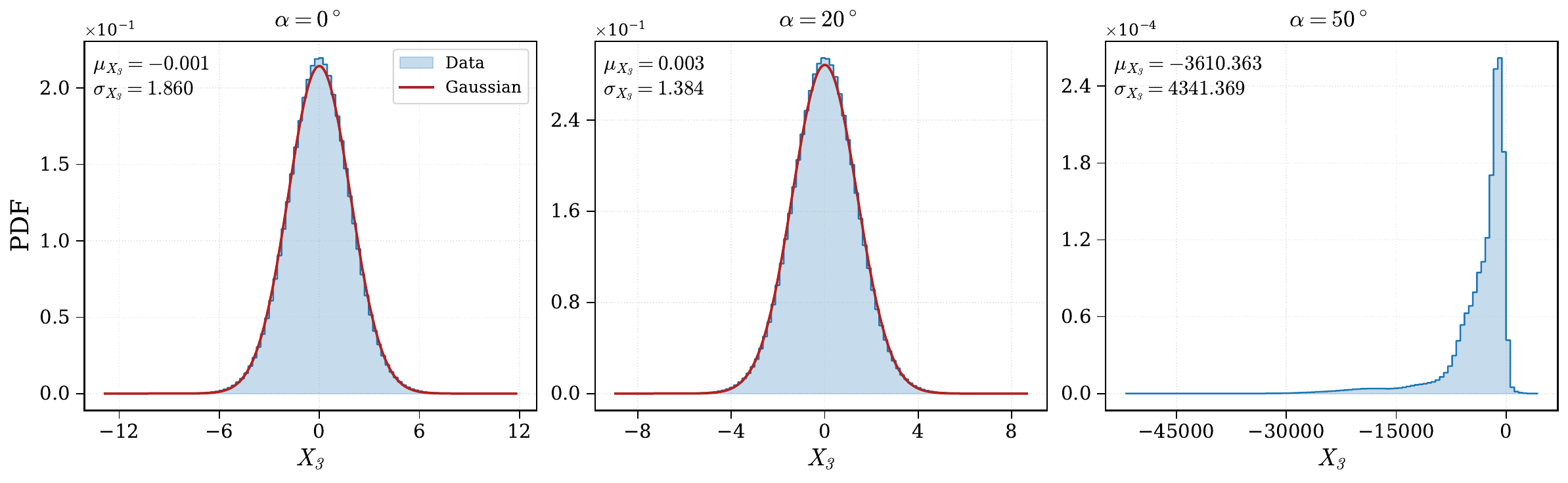}
    \caption{Histograms of the $X_3$ statistic considering all the triangles $(\mathbf k_1,\mathbf k_2,\mathbf k_3)$ whose three sides are within the rectangular region shown in Figure \ref{fig:cylbs_eq}. The three panels represent different PCs ($\alpha=0^{\circ}, 20^{\circ}, \text{and } 50^{\circ}$ ). Solid blue lines denote the data distributions, compared against Gaussian fits shown in red. The values $\mu_{X_3}$ and $\sigma_{X_3}$ denote the mean and standard deviation, respectively, of the $X_3$ distribution for each specific PC.} 
    \label{fig:X3}
\end{figure*}
To assess the statistical properties of the estimated cylindrical bispectrum, we calculate the quantity $X_3$ \citep{gill2026epochreionization21cm}, analogous to the statistic $X$ used by \citet{Pal2022} for the cylindrical power spectrum. $X_3$ is defined as the ratio of $B(k_{1\perp}, k_{2\perp}, k_{3\perp}, k_{1\parallel}, k_{2\parallel})$, the estimated cylindrical BS  to  $\delta {B}_N(k_{1\perp}, k_{2\perp}, k_{3\perp}, k_{1\parallel}, k_{2\parallel})$, the  expected RMS statistical fluctuations arising from system noise, as

\begin{equation}
    X_3 = \dfrac{{B}(k_{1\perp}, k_{2\perp}, k_{3\perp}, k_{1\parallel}, k_{2\parallel})}{\delta {B}_N(k_{1\perp}, k_{2\perp}, k_{3\perp}, k_{1\parallel}, k_{2\parallel})} \,.
    \label{X3}
\end{equation}
We estimate $\delta B_N(k_{1\perp}, k_{2\perp}, k_{3\perp}, k_{1\parallel}, k_{2\parallel})$ from 50 statistically independent realizations of the noise bispectrum $B_N(k_{1\perp}, k_{2\perp}, k_{3\perp}, k_{1\parallel}, k_{2\parallel})$ which were estimated using simulated visibility data that contain only the expected system noise.  
The distribution of $X_3$ is expected to be symmetric about the mean $\mu_{X_3}\approx0$ with standard deviation $\sigma_{X_3} \approx 1$, if the estimated cylindrical BS is consistent with the statistical uncertainties arising from system noise. A nonzero value of $\mu_{X_3}$ or an asymmetric distribution indicates the presence of residual foregrounds or some other systematic effects, while $\sigma_{X_3} > 1$ suggests that the actual fluctuations in the estimated BS are larger than those predicted by our noise-only simulations. Figure~\ref{fig:X3} shows the probability density function (PDF) of $X_3$ for the same three representative PCs as in Figure~\ref{fig:cylbs_eq}. For each PC, we consider $X_3 $ for all the triangles $(\mathbf{k_1},\mathbf{k_2},\mathbf{k_3})$ whose three sides are within the rectangular region shown in \ref{fig:cylbs_eq}. Considering the first two panels, which correspond to $\alpha=0^{\circ}$ and $20^{\circ}$ respectively, the distribution of $X_3$ is nearly symmetric around a mean which has a value close to zero. At $\alpha=0^{\circ}$,  the mean is $\mu_{X_3}= -0.001$ while at $\alpha=20^{\circ}$, the mean is $\mu_{X_3}=0.003$. This indicates that estimates at these two PCs are reasonably free from residual foreground contamination and are primarily dominated by statistical fluctuations. The corresponding standard deviations are $\sigma_{X_3}=1.86$ and $1.384$ respectively, which are also close to unity. This suggests that at these two PCs, the statistical fluctuations in the estimated cylindrical BS  are roughly consistent with the predicted uncertainties due to system noise. In both cases, the $X_3$ distributions closely follow a Gaussian distribution, shown by the red curves in the plots. The behavior changes drastically at the last panel, which corresponds to $\alpha=50^{\circ}$. We see that the PDF is strongly skewed with an extended tail toward the negative side. The mean attains a very large negative value, $\mu_{X_3} \approx -3610.36$, and the standard deviation is also significantly larger, $\sigma_{X_3} \approx 4341.37$. This result is consistent with our previous observation regarding the presence of strong foreground contamination at this PC, particularly due to the very bright source Fornax~A. In the subsequent analysis, to account for the excess statistical fluctuations beyond those predicted by the noise-only simulations, for each PC we scale the noise predictions $\delta B_N$ with the value of $\sigma_{X_3}$ for the corresponding PC, using $\delta B_N^{\rm True} = \delta B_N \times \sigma_{X_3}$. This accounts for the possibility that our noise-only simulations, which use a fixed $T_{sys}=400 \, {\rm K}$ (eq.~1 of \citealt{gill2026epochreionization21cm}),  may underestimate the system noise that is  actually present in our data.

\subsection{Spherical Bispectrum}
\begin{figure*}[t]
    \centering
    \includegraphics[width=\textwidth]{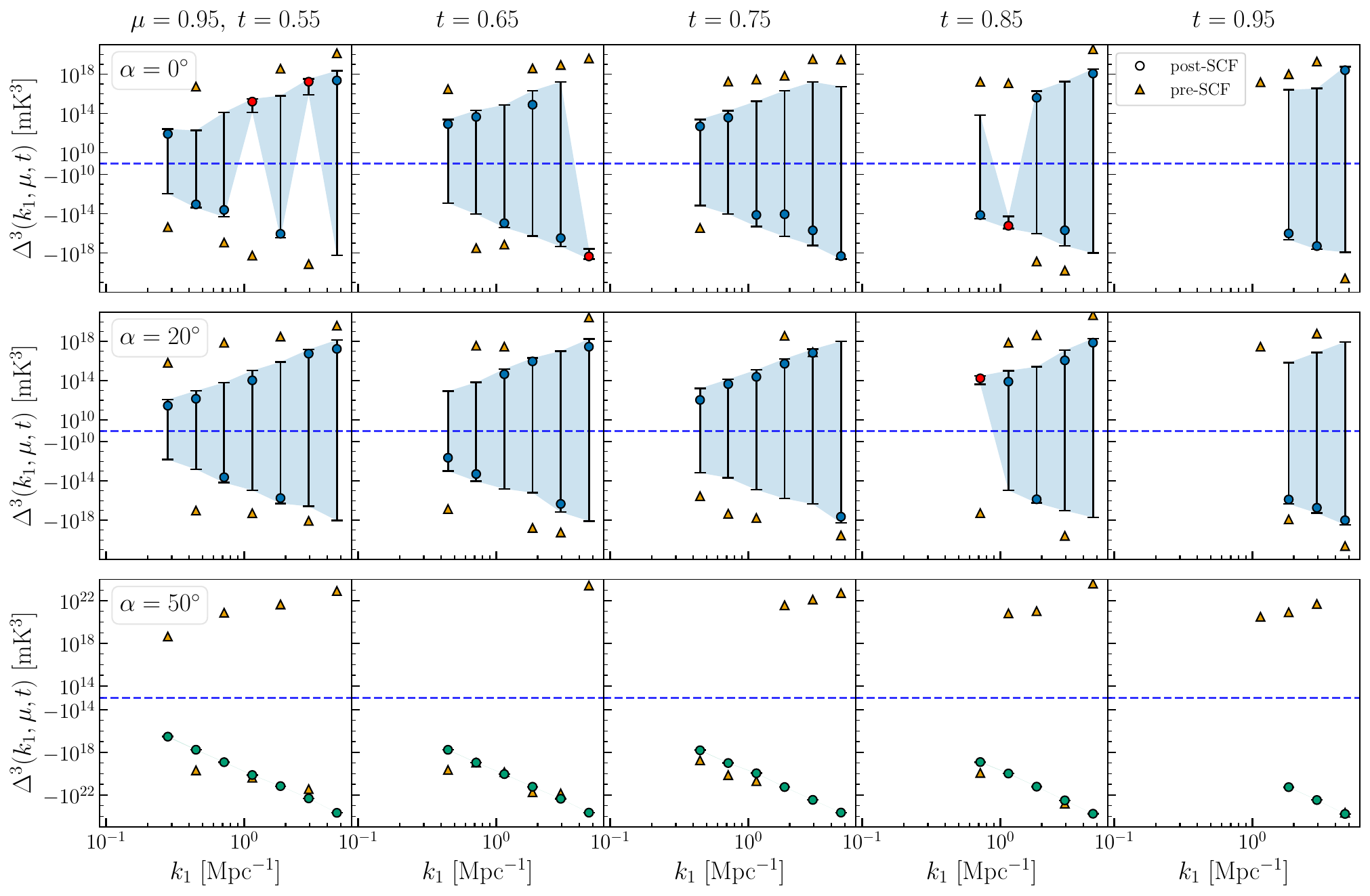}
    \caption{The mean-cubed brightness temperature fluctuation $\Delta^{3}(k_1, \mu, t)$ for nearly linear triangle configurations is plotted as a function of $k_1$. The rows represent different pointing centers (PCs) : $\alpha=0^{\circ}$, $20^{\circ}$, and $50^{\circ}$. The columns correspond to varying triangle shapes defined by the parameters $\mu=0.95$ and $t$ values ranging from $0.55$ to $0.95$. In each panel, orange triangles indicate the results before applying the SCF, whereas colored filled circles represent the post-SCF estimates, with corresponding $2\sigma$ error bars. The shaded regions highlight $2\sigma$ uncertainty region arising from system noise, and the dashed horizontal lines in each panel mark the zero level for the $\Delta^3(k_1,\mu,t)$ magnitude. Estimates within $0\pm2\sigma$ are indicated by blue filled circles, those exceeding $\pm2\sigma$ but remaining within $\pm3\sigma$ by red filled circles, and results inconsistent with fluctuations by green filled circles.}
    \label{fig:3dbs_3pc}
\end{figure*}

In this section we present results for the 3D spherical BS, which is obtained by  spherically binning  the cylindrical BS, as described in Section $\ref{sec:methodology}$, and schematically shown below. 

\[
\resizebox{\columnwidth}{!}{$
\underbrace{
B\left(k_{1\perp},k_{2\perp},k_{3\perp},
k_{1\parallel},k_{2\parallel}\right)
}_{\text{Cylindrical bispectrum}}
\quad
\xrightarrow{\text{Spherical binning}}
\quad
\underbrace{
B(k_1,k_2,k_3)
\equiv B(k_1,\mu,t)
}_{\text{Spherical bispectrum}}
$}
\]
 
  We consider the mean-cubed brightness temperature fluctuations, defined as
\begin{equation}
\Delta^3(k_1,\mu,t) = \frac{k_1^6 \bar{B}(k_1,\mu,t)}{(2\pi^2)^2}.    
\end{equation}
 Considering the entire set of triangles   $(\mathbf{k}_1,\mathbf{k}_2, \mathbf{k}_3) $ that are available in our analysis, we find that the estimated values of $\Delta^3(k_1,\mu,t)$ only cover the $(\mu,t)$ bins where \(\mu = 0.95\), which correspond to nearly linear triangles, as illustrated in  Figure~\ref{fig:ABS_all_shape}. The bin-averaged values of $k_1$, for which estimates are available vary across $(\mu,t)$ bins, spanning the  range $[0.281~{\rm Mpc}^{-1}, 4.699~{\rm Mpc}^{-1}]$. Each panel of  Figure~\ref{fig:3dbs_3pc} shows  $\Delta^3(k_1,\mu,t) $  as a function of \(k_1\), where each column corresponds to a different value of $t$ and 
each row corresponds to a different PC. The three rows correspond to the same three PCs as in Figure~\ref{fig:cylbs_eq}. For comparison, we have also shown the pre-SCF results using the same $k_{\perp}-k_{\parallel}$  region as that used for the post-SCF analysis. The pre-SCF results, which are foreground-dominated, are shown as orange triangles, while the post-SCF results are shown as filled circles. The error-bars show $\pm2\sigma$ statistical fluctuations around the estimated post-SCF $\Delta^3$, where $\sigma=\delta {B}_N^{\rm True}k_1^6/(2\pi^2)^2$. 

The first and second rows show the results for $\alpha = 0^{\circ}$ and $\alpha = 20^{\circ}$, respectively. We observe that both PCs exhibit broadly similar behavior. Before applying SCF, $\Delta^3$ takes both positive and negative values, with amplitude $\sim10^{15}-10^{21}~{\rm mK^3}$, which considerably  exceeds the predicted $2\sigma$ error bars. This indicates that the spherical BS is foreground-dominated before SCF. After SCF, the amplitudes are suppressed by approximately 2--4 orders of magnitude, ranging from $\sim10^{11}$ to $10^{19}~{\rm mK^3}$. Furthermore, most of the estimates of $\Delta^3$ are smaller than the corresponding $2\sigma$ error bars, and are consistent with $2\sigma$ fluctuations around zero. A few estimates are not consistent with $2\sigma$ fluctuations around zero, but lies within the $2\sigma$ and $3\sigma$ range.  Note that the blue filled circles show the values that are within $\pm 2 \sigma$, whereas the red filled circles  exceed  $\pm 2 \sigma$ but are within $\pm 3 \sigma$. 
In contrast, the third row ( $\alpha=50^{\circ}$)  exhibits a completely different behavior. Before applying SCF, $\Delta^3$ exhibits very large values of order $\sim10^{18}-10^{24}~{\rm mK^3}$, which undergo very little suppression and remain of the order of $\sim10^{16} - 10^{24}~{\rm mK^3}$ after SCF.   The $2 \sigma$ error-bars are so small compared to $\Delta^3$, that they are not visible on the scale of the plot, even after SCF. We show these estimates in the last row with green filled circles.  This indicates that for this PC, at $\alpha=50^{\circ}$, our estimates of $\Delta^3$ continue to be dominated by foregrounds despite the application of SCF. This behavior is consistent with our earlier findings from the analysis of the cylindrical BS  and $X_3$.

\subsection{The $\alpha$ Dependence and Constraints on the EoR 21-cm Signal}

\begin{figure*}[htbp]
    \centering
    \includegraphics[width=\textwidth]{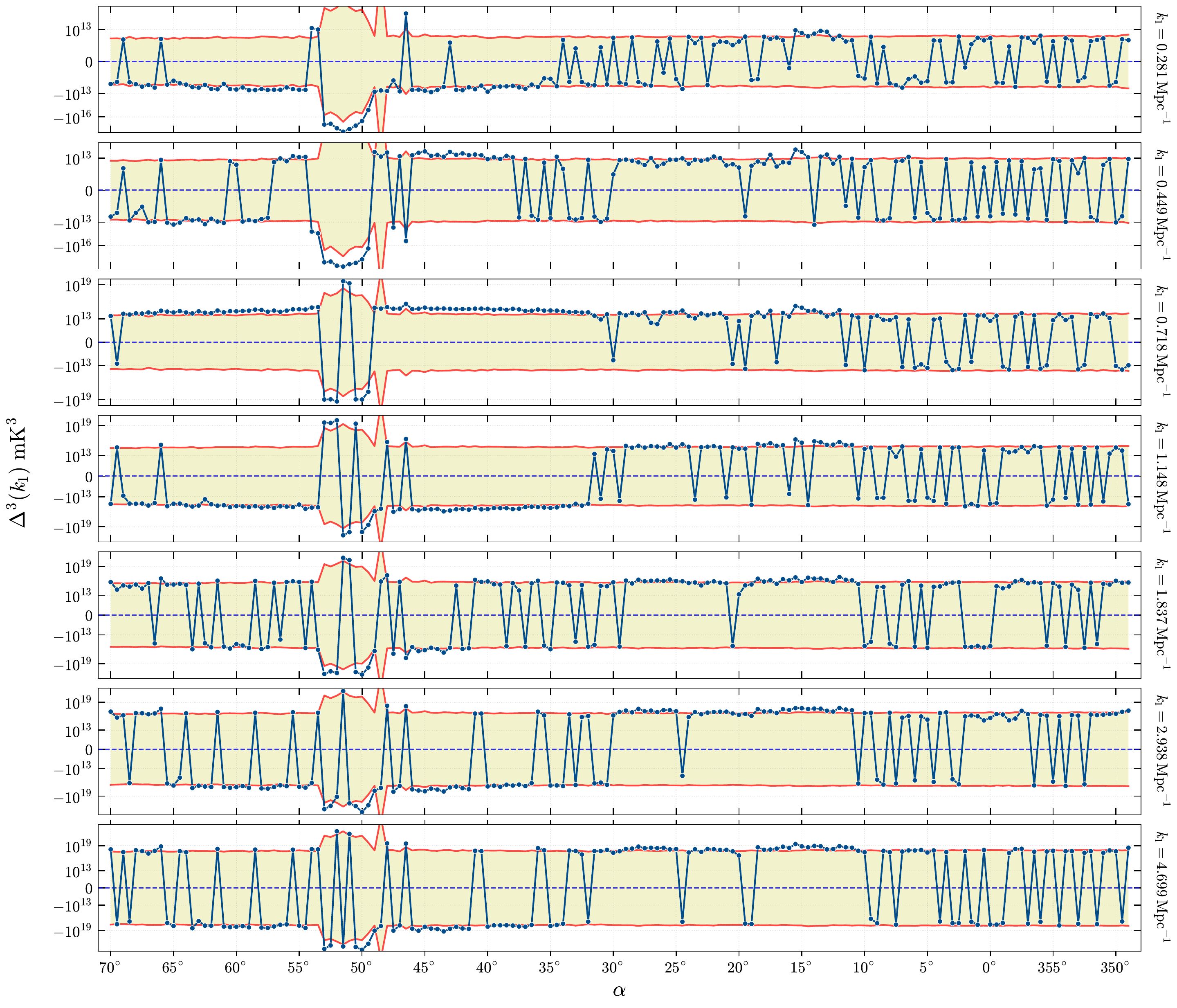}
    \caption{Variation of $\Delta^{3}(k_{1})$ as a function of right ascension $\alpha$. The seven panels correspond to seven $k_1$ bins as indicated at the right side of each panel. The blue solid curve shows the estimated values of $\Delta^{3}$, while the shaded yellow region with red boundary indicates the expected uncertainty arising from system noise around zero. The dashed horizontal line marks the zero level.}
    \label{fig:3dbs_all_k}
\end{figure*}

 In this section, we present  $\Delta^3(k_1)$, after SCF,   as a function of $\alpha$, considering all the PCs included in this drift-scan observation. Each panel  of Figure \ref{fig:3dbs_all_k}  shows $\Delta^3(k_1)$  as a function of $\alpha$ for  a different value of $k_1$. Note that, for every $k_1$  we have collapsed all the $(\mu,t)$ values to obtain $\Delta^3(k_1)$. The results for the full shape-dependent $\Delta^3(k_1,\mu,t)$ are presented in Appendix \ref{appendix:3dbs_kmut}. The yellow shaded region indicates the predicted $2\sigma$ statistical fluctuations around zero.  All the panels exhibit a large negative dip centered at $\alpha \approx 50^{\circ}$, which indicates a large level of residual foreground contamination from Fornax~A. The extent of this contaminated region in $\alpha$ appears to vary across the different $k_1$ bins. For all the $k_1$ bins,  the value of $\Delta^3$ appears to fluctuate randomly within $0 \pm 2 \sigma$ across the $\alpha$ range between $349^{\circ}$ and  $10^{\circ}$, which is roughly consistent with the statistical fluctuations predicted for system noise. The region where $\Delta^3$ is consistent with random fluctuations within  $0 \pm 2 \sigma$ may extend all the way to $\alpha \approx 35^{\circ}$ for the smaller $k_1$ bins; however, this does not hold for the larger $k_1$ bins. In addition, a few PCs in the range $65^{\circ}$--$70^{\circ}$ are consistent with system-noise fluctuations at all the $k_1$ bins.

\begin{table*}[htbp]
\caption{Best limits on the EoR 21-cm bispectrum, at different
$k_1$, after collapsing over all triangle shapes (i.e., all ($\mu$-$t$)
bins). The left half gives the single pointing center (PC) result with the
tightest $2\sigma$ limit, along with the corresponding $\alpha$ and $\Delta^3$. The right half gives the result $\Delta^3_c$ and its $2\sigma_c$ uncertainty,
obtained by incoherently combining the stated number of PCs using
inverse-variance weighting.}
\label{tab:delta3_limits}
\renewcommand{\arraystretch}{1.3}
\begin{tabular}{||c||c|c|c||c|c|c||}
\hline\hline
\multirow{2}{*}{$k_1\,(\mathrm{Mpc}^{-1})$}
& \multicolumn{3}{c||}{Single PC results}
& \multicolumn{3}{c||}{Results after Incoherent Combination} \\
\cline{2-7}
& $\Delta^3 ~({\rm mK^3})$ & $2\sigma~({\rm mK^3})$ & PC
& $\Delta_c^3 ~({\rm mK^3})$ & $2\sigma_c~({\rm mK^3})$ & No. of PCs \\
\hline\hline
0.281 & $(8.00 \times 10^{3})^3$ & $(8.50 \times 10^{3})^3$ & $69^{\circ}$  & $(-4.20 \times 10^{3})^3$ & $(4.89 \times 10^{3})^3$ & 99   \\
\hline
0.449 & $(-8.55 \times 10^{3})^3$ & $(1.61 \times 10^{4})^3$ & $68^{\circ}$ & $(6.64 \times 10^{3})^3$ & $(9.11 \times 10^{3})^3$ & 99 \\
\hline
0.718 & $(-1.68 \times 10^{4})^3$ & $(3.45 \times 10^{4})^3$ & $69.5^{\circ}$ & $(1.97 \times 10^{4})^3$ & $(2.03 \times 10^{4})^3$ & 66  \\
\hline
1.148 & $(-6.34 \times 10^{4})^3$ & $(6.49 \times 10^{4})^3$ & $68^{\circ}$ & $(2.34 \times 10^{4})^3$ & $(3.97 \times 10^{4})^3$ & 75 \\
\hline
1.837 & $(1.17 \times 10^{5})^3$ & $(1.42 \times 10^{5})^3$ & $65.5^{\circ}$ & $(7.92 \times 10^{4})^3$ & $(8.01 \times 10^{4})^3$ & 69 \\
\hline
2.938 & $(-2.49 \times 10^{5})^3$ & $(3.06 \times 10^{5})^3$ & $68.5^{\circ}$ &  $(1.55 \times 10^{5})^3$ & $(1.84 \times 10^{5})^3$ & 66 \\
\hline
4.699 & $(6.66 \times 10^{5})^3$ & $(7.16 \times 10^{5})^3$ & $69^{\circ}$ &  $(3.81 \times 10^{5})^3$ & $(4.35 \times 10^{5})^3$ & 65 \\
\hline\hline
\end{tabular}
\end{table*}

Our observations are not deep enough to detect the expected EoR 21-cm signal, and we interpret the measured $\Delta^3(k_1)$ values as arising from either: (1) statistical fluctuations around zero, provided that  $|\Delta^3(k_1)|\le 2\sigma$; or (2) residual foreground contamination and systematics if $|\Delta^3(k_1)|> 2\sigma$. 
We now only consider the cases where the measured $\Delta^3(k_1)$ is consistent with the predicted statistical fluctuations. We use these measurements to constrain the EoR 21-cm signal such that the value of $\Delta^3(k_1)$ lies within the range  $0\pm 2\sigma(k_1)$.
In the subsequent discussion, we quote the value of $2 \sigma$ to denote both the upper and the lower limit on $\Delta^3(k_1)$, the mean-cubed brightness temperature fluctuations of the EoR 21-cm signal. Considering the different $k_1$ bins  (Figure~\ref{fig:3dbs_all_k}), the left half of Table~\ref{tab:delta3_limits} shows the best value of $2 \sigma$ that is achieved from a single PC. The corresponding values $\Delta^3$  and  $\alpha$ are also shown for reference. The lowest value of $2\sigma$ is obtained in the smallest bin ($k_1 = 0.281~\mathrm{Mpc}^{-1}$) at $\alpha = 69^{\circ}$, yielding $2\sigma = (8.50 \times 10^3)^3~\mathrm{mK}^3$. The value of $2 \sigma$ increases as we move to larger $k$. In all cases (all the $k_1$ bins),  the $\alpha$ values with the lowest $2\sigma$ lie in the range $65^{\circ}$--$70^{\circ}$. 

We next incoherently combine the bispectrum measurements from different PCs in order to tighten the constraints on the EoR 21-cm signal. The results from this analysis are shown in the right half of Table~\ref{tab:delta3_limits}. We consider the inverse-variance-weighted average of 
all the measurements that satisfy  $|\Delta^3(k_1)|\le 2\sigma$ to calculate 
$\Delta^3_c$ and $\sigma_c$,  the corresponding mean and standard deviation respectively. 
We find that the results after incoherent combination satisfy the condition $|\Delta^3_c(k_1)|\le 2\sigma_c$ for all the $k$ bins except $k_1 = 1.837~\mathrm{Mpc}^{-1}$. For this $k$ bin we have excluded an additional 26 PCs  to ensure that the condition $|\Delta^3_c(k_1)|\le 2\sigma_c$ is satisfied. The actual number of PCs that  was used for each $k$ bin is 
listed in Table~\ref{tab:delta3_limits}, and the RA values corresponding to these PCs are tabulated in Appendix~\ref{appendix:table_PCs}.

Considering the results after  incoherent combination (Table~\ref{tab:delta3_limits}), we find the tightest constraint of  $2\sigma_c = (4.89 \times 10^{3})^3~\mathrm{mK}^3$  on the EoR 21-cm signal. This occurs at the lowest $k$ bin with  $k_1 = 0.281~\mathrm{Mpc}^{-1}$, where we have combined $99$ PCs. The number of PCs that satisfy our selection criteria and are combined ranges from $65$ to $99$ across the different $k_1$ bins. We see that the $2 \sigma$ constraint 
tightens by a factor of $\sim 5$ ($4.4-5.6$) across all the $k_1$ bins. As mentioned earlier, the condition $|\Delta^3_c(k_1)|\le 2\sigma_c$ is satisfied in all the $k$ bins; i.e., the results after incoherent combination are all consistent with the statistical fluctuations expected from system noise.

\section{Summary and Conclusions}
\label{sec:sum}
We analyze zenith-pointing  ($\delta = -26.7^\circ$) drift-scan observations at $154.2~{\rm MHz}$ with the Phase~II compact configuration of  MWA. The data span approximately $81^\circ$ in RA, covering the range $349.0^\circ \leq \alpha \leq 70.0^\circ$, with samples taken at intervals of $0.5^\circ$. 
To quantify how the foregrounds for  the EoR 21-cm bispectrum measurements vary with $\alpha$, 
we estimate $\Delta^3_A(\ell_1,\mu_A,t_A)$,
 the 2D mean-cubed brightness temperature fluctuations of the sky signal, for all the pointing centers (PCs) in our drift scan observations. 
For most of the PCs, $|\Delta^3_A|$ increases roughly as $\ell_1^4$, consistent with Poisson fluctuations of extragalactic point sources, and varies only moderately with  $(\mu_A,t_A)$ that quantify the triangle's shape (Figure \ref{fig:ABS_all_shape}). PCs near $\alpha\approx50^\circ$ are an exception: the $\ell_1$ dependence flattens and $|\Delta^3_A|$ at low $\ell_1$ is up to three orders of magnitude larger compared to the other PCs. The $\alpha$ dependence of $|\Delta^3_A|$ 
shows a dominant peak near $\alpha\approx50^\circ$ at low $\ell_1$, (Figure \ref{fig:ABS_heatmap}), which we attribute to the presence of a bright extended source, Fornax~A, in the primary beam. This $\alpha$ dependence weakens at larger $\ell_1$, where the signal becomes increasingly dominated by point-source Poisson fluctuations. The region $349^\circ\lesssim\alpha\lesssim30^\circ$ is comparatively less affected by foregrounds at low $\ell_1$ and is therefore the most promising for constraining the EoR 21-cm bispectrum. The results presented here are very similar to those in \citet{sarkar_2026} who have analyzed the same drift-scan observations to quantify how the foregrounds for the EoR 21-cm power spectrum measurements vary with $\alpha$.

We  have estimated the 3D bispectrum (BS) of the sky signal for all the PCs in our drift scan observations,  after applying SCF to mitigate the foreground contamination. We analyze $B(k_{1\perp}, k_{2\perp}, k_{3\perp}, k_{1\parallel}, k_{2\parallel})$, the cylindrical BS (Figure~\ref{fig:cylbs_eq}) and the $X_3$ statistics (Figure~\ref{fig:X3}) to assess the residual foreground contamination after SCF. Considering  PCs located far away from Fornax~A, we find that the triangles where $(\mathbf{k_1},\mathbf{k_2},\mathbf{k_3})$ are restricted within the region $k_{\parallel}\ge0.135\,\mathrm{Mpc}^{-1}$ and $k_{\perp}\leq0.026\,\mathrm{Mpc}^{-1}$ are relatively free from foreground contamination, in contrast to the PCs near Fornax~A where the entire $(k_{\perp},k_{\parallel})$ space is contaminated by residual foregrounds. This is borne out by the  $X_3$ values, which have a symmetric Gaussian distribution with $\mu_{X_3} \approx 0$ and $\sigma_{X_3} \approx 1$ for the clean $(k_{\perp},k_{\parallel})$ regions. In contrast,  $X_3$ exhibits a highly asymmetric  distribution around a large negative $\mu_{X_3}$ with $\sigma_{X_3} \gg 1$ in the same $(k_{\perp},k_{\parallel})$ region for the contaminated PCs. 

We have used all the triangles $(\mathbf{k_1},\mathbf{k_2},\mathbf{k_3})$ that lie within the region $k_{\parallel}\ge0.135\,\mathrm{Mpc}^{-1}$ and  $k_{\perp}\leq0.026\,\mathrm{Mpc}^{-1}$ to calculate the spherical BS for each PC. Figure~\ref{fig:3dbs_all_k} shows $\Delta^3(k)$, the mean-cubed brightness temperature fluctuations, as a function of $\alpha$ for all the $k_1$ bins after collapsing all the triangle shapes.   All the $k_1$ bins show large negative values of $\Delta^3$ in the proximity of $\alpha=50^{\circ}$, indicating a large foreground contamination from Fornax~A.  Away from the contaminated $\alpha$ range, whose extent varies with $k$ bin, the measured values of $\Delta^3(k)$ fluctuate around zero, consistent with the statistical fluctuations expected due to system noise   $(|\Delta^3|  \le 2 \sigma)$. Considering all the  $k$ bins, this condition is satisfied for most of the PCs in the range $\alpha=349^{\circ}$ to $\approx 10^{\circ}$, and also some of the PCs in the range $65^{\circ}$--$70^{\circ}$. We use the  PCs where the condition $| \Delta^3 | \le 2 \sigma$ is satisfied to constrain the value of $\Delta^3(k)$ for the EoR 21-cm signal to lie in the range $0 \pm 2 \sigma$. Here, we quote the value of $2 \sigma$  to denote both the upper limit $(2 \sigma)$ and the lower limit $(-2 \sigma)$ on  $\Delta^3(k)$ for the EoR 21-cm signal. Considering all the PCs, the best limit of $2\sigma=(8.50\times10^3)^3~{\rm mK^3}$ is achieved at $\alpha=69^{\circ}$ for the lowest bin $k_1=0.281~{\rm Mpc}^{-1}$.

In order to further tighten the constraints on the EoR 21-cm bispectrum, we have incoherently combined the values of  $\Delta^3$ obtained from multiple PCs which satisfy the condition $|\Delta^3(k_1)| \le 2\sigma$  to obtain $\Delta^3_c$ and $\sigma_c$, the corresponding mean and standard deviation, respectively. Considering the smallest bin at $k_1=0.281~{\rm Mpc}^{-1}$, we are able to combine 99 PCs to obtain the tightest constraint of $2\sigma_c=(4.89\times10^3)^3~{\rm mK^3}$. This represents a substantial improvement, as this is a factor of $5.25$ lower than the best single-PC limit.  Furthermore, this limit is $\sim 16$ times tighter than the previous constraint of $2\sigma = (1.24 \times 10^4)^3~{\rm mK^3}$ reported by \citet{gill2026epochreionization21cm}. To the best of our knowledge, the results reported in this work represent the tightest constraint on the EoR 21-cm bispectrum to date.

 For completeness, we note the study by  \citet{sarkar_2026} who have analyzed the same data to constrain the EoR 21-cm PS and study this as a function of $\alpha$. They were able to incoherently combine $23$ PCs to obtain a $2\sigma$ upper limit of $\Delta^2_{\rm UL}=(98.67)^2{\rm mK}^2$ at $k=0.156~{\rm Mpc}^{-1}$, which represents the best upper limit on the mean squared brightness temperature fluctuations of the EoR 21-cm signal at $z = 8.2$ from MWA. Their upper limit remains $\approx 3$ orders of magnitude higher than the 
 predicted 
 EoR 21-cm signal $\Delta^2\sim 10~{\rm mK}^2$ 
 \citep{mondal_2017}. The mean-cubed brightness-temperature fluctuation predicted for the EoR 21-cm signal at $z \approx 8$ is $|\Delta^3| \approx 10^3~{\rm mK}^3$ at $k_1 \approx 0.29~{\rm Mpc}^{-1}$ \citep{gill_eormulti}. Although we are able to incoherently combine $99$ PCs,   our limits are still around eight orders of magnitude higher than the predicted signal. 

 Finally, we note that incoherent combination of multiple PCs causes the statistical fluctuations due to noise to scale as $\sim t^{-1/2}$for both the PS and the BS, where $t$ is the total observing time of the observation.   In contrast, we expect a $t^{-1}$ and a $t^{-1.5}$ scaling for the PS and BS, respectively, in case of targeted observations or coherent combination of multiple PCs.  It is quite clear that either deep targeted observations or the coherent combination of multiple PCs will lead to a substantial gain, particularly for the BS. Efforts are currently underway to develop a coherent combination strategy. Furthermore, we are currently unable to access the low $k$ values owing to the signal loss due to SCF. In future work, we  plan to investigate foreground mitigation using 
 Bayes-SCF \citep{elahi_2026}, which might allow us to access smaller $k$ values. 
  Our study is expected to provide useful guidance for selecting target fields for future EoR 21-cm observations with the Square Kilometer Array (SKA) which is at the same location as MWA, and which will see the same region of the sky as analyzed here. 
  
\section*{Acknowledgments}
 We acknowledge the National Supercomputing Mission (NSM) for providing computing resources of ‘PARAM Shakti’ at IIT Kharagpur, which is implemented by C-DAC and supported by the Ministry of Electronics and Information Technology (MeitY) and the Department of Science and Technology (DST), Government of India. We also acknowledge the computing facilities provided by the Department of Physics at IIT Kharagpur.

\section*{Data Availability}

The data sets were derived from sources in the public domain
(the MWA Data Archive: project ID G0031) available at \href{https://asvo.mwatelescope.org/}{MWA ASVO Portal}.


\bibliography{main}{}

\begin{thebibliography}{}
\expandafter\ifx\csname natexlab\endcsname\relax\def\natexlab#1{#1}\fi
\providecommand{\url}[1]{\href{#1}{#1}}
\providecommand{\dodoi}[1]{doi:~\href{http://doi.org/#1}{\nolinkurl{#1}}}
\providecommand{\doeprint}[1]{\href{http://ascl.net/#1}{\nolinkurl{http://ascl.net/#1}}}
\providecommand{\doarXiv}[1]{\href{https://arxiv.org/abs/#1}{\nolinkurl{https://arxiv.org/abs/#1}}}

\bibitem[{Ali {et~al.}(2008)Ali, Bharadwaj, \& Chengalur}]{ali}
Ali, S.~S., Bharadwaj, S., \& Chengalur, J.~N. 2008, Monthly Notices of the Royal Astronomical Society, 385, 2166, \dodoi{10.1111/j.1365-2966.2008.12984.x}

\bibitem[{{Bernardi} {et~al.}(2009){Bernardi}, {de Bruyn}, {Brentjens}, {Ciardi}, {Harker}, {Jeli{\'c}}, {Koopmans}, {Labropoulos}, {Offringa}, {Pandey}, {Schaye}, {Thomas}, {Yatawatta}, \& {Zaroubi}}]{Bernardi2009}
{Bernardi}, G., {de Bruyn}, A.~G., {Brentjens}, M.~A., {et~al.} 2009, \aap, 500, 965, \dodoi{10.1051/0004-6361/200911627}

\bibitem[{{Bharadwaj} {et~al.}(2020){Bharadwaj}, {Mazumdar}, \& {Sarkar}}]{bharadwaj_2020}
{Bharadwaj}, S., {Mazumdar}, A., \& {Sarkar}, D. 2020, \mnras, 493, 594, \dodoi{10.1093/mnras/staa279}

\bibitem[{Bharadwaj \& Pandey(2005)}]{BP2005}
Bharadwaj, S., \& Pandey, S.~K. 2005, Mon. Not. Roy. Astron. Soc., 358, 968, \dodoi{10.1111/j.1365-2966.2005.08836.x}

\bibitem[{Carroll {et~al.}(2016)Carroll, Line, Morales, Barry, Beardsley, Hazelton, Jacobs, Pober, Sullivan, Webster, Bernardi, Bowman, Briggs, Cappallo, Corey, de~Oliveira-Costa, Dillon, Emrich, Ewall-Wice, Feng, Gaensler, Goeke, Greenhill, Hewitt, Hurley-Walker, Johnston-Hollitt, Kaplan, Kasper, Kim, Kratzenberg, Lenc, Loeb, Lonsdale, Lynch, McKinley, McWhirter, Mitchell, Morgan, Neben, Oberoi, Offringa, Ord, Paul, Pindor, Prabu, Procopio, Riding, Rogers, Roshi, Shankar, Sethi, Srivani, Subrahmanyan, Tegmark, Thyagarajan, Tingay, Trott, Waterson, Wayth, Whitney, Williams, Williams, Wu, \& Wyithe}]{Carroll2016}
Carroll, P.~A., Line, J., Morales, M.~F., {et~al.} 2016, Monthly Notices of the Royal Astronomical Society, 461, 4151, \dodoi{10.1093/mnras/stw1599}

\bibitem[{{Chatterjee} {et~al.}(2024){Chatterjee}, {Elahi}, {Bharadwaj}, {Sarkar}, {Choudhuri}, {Sethi}, \& {Patwa}}]{chatterjee_2024}
{Chatterjee}, S., {Elahi}, K. M.~A., {Bharadwaj}, S., {et~al.} 2024, \pasa, 41, e077, \dodoi{10.1017/pasa.2024.45}

\bibitem[{Chatterjee {et~al.}(2025)Chatterjee, Sarkar, Choudhuri, Elahi, Bharadwaj, Sethi, \& Patwa}]{Chatterjee_2025}
Chatterjee, S., Sarkar, S., Choudhuri, S., {et~al.} 2025, Publications of the Astronomical Society of Australia, 42, e103, \dodoi{10.1017/pasa.2025.10065}

\bibitem[{{DeBoer} {et~al.}(2017){DeBoer}, {Parsons}, {Aguirre}, {Alexander}, {Ali}, {Beardsley}, {Bernardi}, {Bowman}, {Bradley}, {Carilli}, {Cheng}, {de Lera Acedo}, {Dillon}, {Ewall-Wice}, {Fadana}, {Fagnoni}, {Fritz}, {Furlanetto}, {Glendenning}, {Greig}, {Grobbelaar}, {Hazelton}, {Hewitt}, {Hickish}, {Jacobs}, {Julius}, {Kariseb}, {Kohn}, {Lekalake}, {Liu}, {Loots}, {MacMahon}, {Malan}, {Malgas}, {Maree}, {Martinot}, {Mathison}, {Matsetela}, {Mesinger}, {Morales}, {Neben}, {Patra}, {Pieterse}, {Pober}, {Razavi-Ghods}, {Ringuette}, {Robnett}, {Rosie}, {Sell}, {Smith}, {Syce}, {Tegmark}, {Thyagarajan}, {Williams}, \& {Zheng}}]{DeBoer_2017}
{DeBoer}, D.~R., {Parsons}, A.~R., {Aguirre}, J.~E., {et~al.} 2017, \pasp, 129, 045001, \dodoi{10.1088/1538-3873/129/974/045001}

\bibitem[{Elahi(2026)}]{elahi_2026}
Elahi, K. M.~A. 2026, Monthly Notices of the Royal Astronomical Society, 550, stag1225, \dodoi{10.1093/mnras/stag1225}

\bibitem[{{Elahi} {et~al.}(2025){Elahi}, {Bharadwaj}, {Chatterjee}, {Sarkar}, {Choudhuri}, {Sethi}, \& {Patwa}}]{Elahi_missing}
{Elahi}, K. M.~A., {Bharadwaj}, S., {Chatterjee}, S., {et~al.} 2025, \mnras, 540, 2745, \dodoi{10.1093/mnras/staf896}

\bibitem[{{EoR/CD Science Working Group} {et~al.}(2026){EoR/CD Science Working Group}, {Bernardi}, {Breitman}, {Datta}, {Fialkov}, {Koopmans}, {Liu}, {Mao}, {Mellema}, {Mertens}, \& et~al.}]{EoR_SWG_arXiV_2026}
{EoR/CD Science Working Group}, {Bernardi}, G., {Breitman}, D., {et~al.} 2026, arXiv e-prints, arXiv:2606.26435, \dodoi{10.48550/arXiv.2606.26435}

\bibitem[{{Ghosh} {et~al.}(2012){Ghosh}, {Prasad}, {Bharadwaj}, {Ali}, \& {Chengalur}}]{Ghosh2012}
{Ghosh}, A., {Prasad}, J., {Bharadwaj}, S., {Ali}, S.~S., \& {Chengalur}, J.~N. 2012, \mnras, 426, 3295, \dodoi{10.1111/j.1365-2966.2012.21889.x}

\bibitem[{Gill \& Bharadwaj(2025)}]{Gill_2025_mabs}
Gill, S.~S., \& Bharadwaj, S. 2025, The Astrophysical Journal, 995, 175, \dodoi{10.3847/1538-4357/ae160c}

\bibitem[{{Gill} {et~al.}(2025{\natexlab{a}}){Gill}, {Bharadwaj}, {Ali}, \& {Elahi}}]{Gill_2024_2d3vc}
{Gill}, S.~S., {Bharadwaj}, S., {Ali}, S.~S., \& {Elahi}, K. M.~A. 2025{\natexlab{a}}, \apj, 979, 25, \dodoi{10.3847/1538-4357/ad9b20}

\bibitem[{{Gill} {et~al.}(2025{\natexlab{b}}){Gill}, {Bharadwaj}, {Elahi}, {Sethi}, \& {Patwa}}]{Gill_2025_mwa1}
{Gill}, S.~S., {Bharadwaj}, S., {Elahi}, K. M.~A., {Sethi}, S.~K., \& {Patwa}, A.~K. 2025{\natexlab{b}}, \apj, 993, 56, \dodoi{10.3847/1538-4357/ae0463}

\bibitem[{{Gill} {et~al.}(2026){Gill}, {Bharadwaj}, {Elahi}, {Sethi}, \& {Patwa}}]{gill2026epochreionization21cm}
---. 2026, The Epoch of Reionization 21 cm Bispectrum at $z=8.2$ from MWA data II: Smooth Component Filtering.
\newblock \doarXiv{2602.17304}

\bibitem[{{Gill} {et~al.}(2024){Gill}, {Pramanick}, {Bharadwaj}, {Shaw}, \& {Majumdar}}]{gill_eormulti}
{Gill}, S.~S., {Pramanick}, S., {Bharadwaj}, S., {Shaw}, A.~K., \& {Majumdar}, S. 2024, \mnras, 527, 1135, \dodoi{10.1093/mnras/stad3273}

\bibitem[{{Gupta} {et~al.}(2017){Gupta}, {Ajithkumar}, {Kale}, {Nayak}, {Sabhapathy}, {Sureshkumar}, {Swami}, {Chengalur}, {Ghosh}, {Ishwara-Chandra}, {Joshi}, {Kanekar}, {Lal}, \& {Roy}}]{Gupta_2017}
{Gupta}, Y., {Ajithkumar}, B., {Kale}, H.~S., {et~al.} 2017, Current Science, 113, 707, \dodoi{10.18520/cs/v113/i04/707-714}

\bibitem[{{HERA Collaboration} {et~al.}(2023){HERA Collaboration}, {Abdurashidova}, {Adams}, {Aguirre}, {Alexander}, {Ali}, {Baartman}, {Balfour}, {Barkana}, {Beardsley}, {Bernardi}, {Billings}, {Bowman}, {Bradley}, {Breitman}, {Bull}, {Burba}, {Carey}, {Carilli}, {Cheng}, {Choudhuri}, {DeBoer}, {de Lera Acedo}, {Dexter}, {Dillon}, {Ely}, {Ewall-Wice}, {Fagnoni}, {Fialkov}, {Fritz}, {Furlanetto}, {Gale-Sides}, {Garsden}, {Glendenning}, {Gorce}, {Gorthi}, {Greig}, {Grobbelaar}, {Halday}, {Hazelton}, {Heimersheim}, {Hewitt}, {Hickish}, {Jacobs}, {Julius}, {Kern}, {Kerrigan}, {Kittiwisit}, {Kohn}, {Kolopanis}, {Lanman}, {La Plante}, {Lewis}, {Liu}, {Loots}, {Ma}, {MacMahon}, {Malan}, {Malgas}, {Malgas}, {Maree}, {Marero}, {Martinot}, {McBride}, {Mesinger}, {Mirocha}, {Molewa}, {Morales}, {Mosiane}, {Mu{\~n}oz}, {Murray}, {Nagpal}, {Neben}, {Nikolic}, {Nunhokee}, {Nuwegeld}, {Parsons}, {Pascua}, {Patra}, {Pieterse}, {Qin}, {Razavi-Ghods}, {Robnett}, {Rosie}, {Santos}, {Sims}, {Singh}, {Smith}, {Swarts}, {Tan},
  {Thyagarajan}, {Wilensky}, {Williams}, {van Wyngaarden}, \& {Zheng}}]{HERA2023}
{HERA Collaboration}, {Abdurashidova}, Z., {Adams}, T., {et~al.} 2023, \apj, 945, 124, \dodoi{10.3847/1538-4357/acaf50}

\bibitem[{Hutter {et~al.}(2020)Hutter, Watkinson, Seiler, Dayal, Sinha, \& Croton}]{Hutter_2019}
Hutter, A., Watkinson, C.~A., Seiler, J., {et~al.} 2020, MNRAS, 492, 653, \dodoi{10.1093/mnras/stz3139}

\bibitem[{{Koopmans} {et~al.}(2015){Koopmans}, {Pritchard}, {Mellema}, {Aguirre}, {Ahn}, {Barkana}, {van Bemmel}, {Bernardi}, {Bonaldi}, {Briggs}, {de Bruyn}, {Chang}, {Chapman}, {Chen}, {Ciardi}, {Dayal}, {Ferrara}, {Fialkov}, {Fiore}, {Ichiki}, {Illiev}, {Inoue}, {Jelic}, {Jones}, {Lazio}, {Maio}, {Majumdar}, {Mack}, {Mesinger}, {Morales}, {Parsons}, {Pen}, {Santos}, {Schneider}, {Semelin}, {de Souza}, {Subrahmanyan}, {Takeuchi}, {Vedantham}, {Wagg}, {Webster}, {Wyithe}, {Datta}, \& {Trott}}]{Koopmans_2015}
{Koopmans}, L., {Pritchard}, J., {Mellema}, G., {et~al.} 2015, in Advancing Astrophysics with the Square Kilometre Array (AASKA14), 1, \dodoi{10.22323/1.215.0001}

\bibitem[{Majumdar {et~al.}(2020)Majumdar, Kamran, Pritchard, Mondal, Mazumdar, Bharadwaj, \& Mellema}]{Majumdar_2020}
Majumdar, S., Kamran, M., Pritchard, J.~R., {et~al.} 2020, MNRAS, 499, 5090, \dodoi{10.1093/mnras/staa3168}

\bibitem[{Majumdar {et~al.}(2018)Majumdar, Pritchard, Mondal, Watkinson, Bharadwaj, \& Mellema}]{Majumdar_2018}
Majumdar, S., Pritchard, J.~R., Mondal, R., {et~al.} 2018, MNRAS, 476, 4007, \dodoi{10.1093/mnras/sty535}

\bibitem[{McKinley {et~al.}(2015)McKinley, Yang, López-Caniego, Briggs, Hurley-Walker, Wayth, Offringa, Crocker, Bernardi, Procopio, Gaensler, Tingay, Johnston-Hollitt, McDonald, Bell, Bhat, Bowman, Cappallo, Corey, Deshpande, Emrich, Ewall-Wice, Feng, Goeke, Greenhill, Hazelton, Hewitt, Hindson, Jacobs, Kaplan, Kasper, Kratzenberg, Kudryavtseva, Lenc, Lonsdale, Lynch, McWhirter, Mitchell, Morales, Morgan, Oberoi, Ord, Pindor, Prabu, Riding, Rogers, Roshi, Udaya~Shankar, Srivani, Subrahmanyan, Waterson, Webster, Whitney, Williams, \& Williams}]{McKinley2025}
McKinley, B., Yang, R., López-Caniego, M., {et~al.} 2015, Monthly Notices of the Royal Astronomical Society, 446, 3478, \dodoi{10.1093/mnras/stu2310}

\bibitem[{Mondal {et~al.}(2017)Mondal, Bharadwaj, \& Majumdar}]{mondal_2017}
Mondal, R., Bharadwaj, S., \& Majumdar, S. 2017, Monthly Notices of the Royal Astronomical Society, 464, 2992, \dodoi{10.1093/mnras/stw2599}

\bibitem[{{Mondal} {et~al.}(2015){Mondal}, {Bharadwaj}, {Majumdar}, {Bera}, \& {Acharyya}}]{mondal2015}
{Mondal}, R., {Bharadwaj}, S., {Majumdar}, S., {Bera}, A., \& {Acharyya}, A. 2015, \mnras, 449, L41, \dodoi{10.1093/mnrasl/slv015}

\bibitem[{Murmu {et~al.}(2023)Murmu, Datta, Majumdar, \& Greve}]{Murmu:2023fgv}
Murmu, C.~S., Datta, K.~K., Majumdar, S., \& Greve, T.~R. 2023.
\newblock \doarXiv{2311.17062}

\bibitem[{{Nunhokee} {et~al.}(2025){Nunhokee}, {Null}, {Trott}, {Barry}, {Qin}, {Wayth}, {Line}, {Jordan}, {Pindor}, {Cook}, {Bowman}, {Chokshi}, {Ducharme}, {Elder}, {Guo}, {Hazelton}, {Hidayat}, {Ito}, {Jacobs}, {Jong}, {Kolopanis}, {Kunicki}, {Lilleskov}, {Morales}, {Pober}, {Selvaraj}, {Shi}, {Takahashi}, {Tingay}, {Webster}, {Yoshiura}, \& {Zheng}}]{Nunhokee2025}
{Nunhokee}, C.~D., {Null}, D., {Trott}, C.~M., {et~al.} 2025, \apj, 989, 57, \dodoi{10.3847/1538-4357/adda45}

\bibitem[{{Nuttall}(1981)}]{Nuttall_1981}
{Nuttall}, A.~H. 1981, IEEE Transactions on Acoustics Speech and Signal Processing, 29, 84

\bibitem[{{Pal} {et~al.}(2022){Pal}, {Elahi}, {Bharadwaj}, {Ali}, {Choudhuri}, {Ghosh}, {Chakraborty}, {Datta}, {Roy}, {Choudhury}, \& {Dutta}}]{Pal2022}
{Pal}, S., {Elahi}, K. M.~A., {Bharadwaj}, S., {et~al.} 2022, \mnras, 516, 2851, \dodoi{10.1093/mnras/stac2419}

\bibitem[{{Patwa} {et~al.}(2021){Patwa}, {Sethi}, \& {Dwarakanath}}]{Patwa_2021}
{Patwa}, A.~K., {Sethi}, S., \& {Dwarakanath}, K.~S. 2021, \mnras, 504, 2062, \dodoi{10.1093/mnras/stab989}

\bibitem[{{Planck Collaboration} {et~al.}(2020){Planck Collaboration}, {Akrami}, {Arroja}, {Ashdown}, {Aumont}, {Baccigalupi}, {Ballardini}, {Banday}, {Barreiro}, {Bartolo}, {Basak}, {Benabed}, {Bernard}, {Bersanelli}, {Bielewicz}, {Bond}, {Borrill}, {Bouchet}, {Bucher}, {Burigana}, {Butler}, {Calabrese}, {Cardoso}, {Casaponsa}, {Challinor}, {Chiang}, {Colombo}, {Combet}, {Crill}, {Cuttaia}, {de Bernardis}, {de Rosa}, {de Zotti}, {Delabrouille}, {Delouis}, {Di Valentino}, {Diego}, {Dor{\'e}}, {Douspis}, {Ducout}, {Dupac}, {Dusini}, {Efstathiou}, {Elsner}, {En{\ss}lin}, {Eriksen}, {Fantaye}, {Fergusson}, {Fernandez-Cobos}, {Finelli}, {Frailis}, {Fraisse}, {Franceschi}, {Frolov}, {Galeotta}, {Galli}, {Ganga}, {G{\'e}nova-Santos}, {Gerbino}, {Gonz{\'a}lez-Nuevo}, {G{\'o}rski}, {Gratton}, {Gruppuso}, {Gudmundsson}, {Hamann}, {Handley}, {Hansen}, {Herranz}, {Hivon}, {Huang}, {Jaffe}, {Jones}, {Jung}, {Keih{\"a}nen}, {Keskitalo}, {Kiiveri}, {Kim}, {Krachmalnicoff}, {Kunz}, {Kurki-Suonio}, {Lamarre}, {Lasenby},
  {Lattanzi}, {Lawrence}, {Le Jeune}, {Levrier}, {Lewis}, {Liguori}, {Lilje}, {Lindholm}, {L{\'o}pez-Caniego}, {Ma}, {Mac{\'\i}as-P{\'e}rez}, {Maggio}, {Maino}, {Mandolesi}, {Marcos-Caballero}, {Maris}, {Martin}, {Mart{\'\i}nez-Gonz{\'a}lez}, {Matarrese}, {Mauri}, {McEwen}, {Meerburg}, {Meinhold}, {Melchiorri}, {Mennella}, {Migliaccio}, {Miville-Desch{\^e}nes}, {Molinari}, {Moneti}, {Montier}, {Morgante}, {Moss}, {M{\"u}nchmeyer}, {Natoli}, {Oppizzi}, {Pagano}, {Paoletti}, {Partridge}, {Patanchon}, {Perrotta}, {Pettorino}, {Piacentini}, {Polenta}, {Puget}, {Rachen}, {Racine}, {Reinecke}, {Remazeilles}, {Renzi}, {Rocha}, {Rubi{\~n}o-Mart{\'\i}n}, {Ruiz-Granados}, {Salvati}, {Savelainen}, {Scott}, {Shellard}, {Shiraishi}, {Sirignano}, {Sirri}, {Smith}, {Spencer}, {Stanco}, {Sunyaev}, {Suur-Uski}, {Tauber}, {Tavagnacco}, {Tenti}, {Toffolatti}, {Tomasi}, {Trombetti}, {Valiviita}, {Van Tent}, {Vielva}, {Villa}, {Vittorio}, {Wandelt}, {Wehus}, {Zacchei}, \& {Zonca}}]{planck2020}
{Planck Collaboration}, {Akrami}, Y., {Arroja}, F., {et~al.} 2020, \aap, 641, A9, \dodoi{10.1051/0004-6361/201935891}

\bibitem[{{Prabu} {et~al.}(2015){Prabu}, {Srivani}, {Roshi}, {Kamini}, {Madhavi}, {Emrich}, {Crosse}, {Williams}, {Waterson}, {Deshpande}, {Shankar}, {Subrahmanyan}, {Briggs}, {Goeke}, {Tingay}, {Johnston-Hollitt}, {R}, {Morgan}, {Pathikulangara}, {Bunton}, {Hampson}, {Williams}, {Ord}, {Wayth}, {Kumar}, {Morales}, {deSouza}, {Kratzenberg}, {Pallot}, {McWhirter}, {Hazelton}, {Arcus}, {Barnes}, {Bernardi}, {Booler}, {Bowman}, {Cappallo}, {Corey}, {Greenhill}, {Herne}, {Hewitt}, {Kaplan}, {Kasper}, {Kincaid}, {Koenig}, {Lonsdale}, {Lynch}, {Mitchell}, {Oberoi}, {Remillard}, {Rogers}, {Salah}, {Sault}, {Stevens}, {Tremblay}, {Webster}, {Whitney}, \& {Wyithe}}]{Prabu2015}
{Prabu}, T., {Srivani}, K.~S., {Roshi}, D.~A., {et~al.} 2015, Experimental Astronomy, 39, 73, \dodoi{10.1007/s10686-015-9444-3}

\bibitem[{{Raste} {et~al.}(2024){Raste}, {Kulkarni}, {Watkinson}, {Keating}, \& {Haehnelt}}]{raste_2023}
{Raste}, J., {Kulkarni}, G., {Watkinson}, C.~A., {Keating}, L.~C., \& {Haehnelt}, M.~G. 2024, \mnras, 529, 129, \dodoi{10.1093/mnras/stae492}

\bibitem[{Sarkar {et~al.}(2026)Sarkar, Elahi, Choudhuri, Bharadwaj, Chatterjee, Bhattacharyya, Sethi, \& Patwa}]{sarkar_2026}
Sarkar, S., Elahi, K. M.~A., Choudhuri, S., {et~al.} 2026, Monthly Notices of the Royal Astronomical Society, stag801, \dodoi{10.1093/mnras/stag801}

\bibitem[{Saxena {et~al.}(2020)Saxena, Majumdar, Kamran, \& Viel}]{Saxena_2020}
Saxena, A., Majumdar, S., Kamran, M., \& Viel, M. 2020, Mon. Not. Roy. Astron. Soc., 497, 2941, \dodoi{10.1093/mnras/staa1768}

\bibitem[{{Shimabukuro} {et~al.}(2016){Shimabukuro}, {Yoshiura}, {Takahashi}, {Yokoyama}, \& {Ichiki}}]{Shimabukuro_2016}
{Shimabukuro}, H., {Yoshiura}, S., {Takahashi}, K., {Yokoyama}, S., \& {Ichiki}, K. 2016, \mnras, 458, 3003, \dodoi{10.1093/mnras/stw482}

\bibitem[{{Swarup} {et~al.}(1991){Swarup}, {Ananthakrishnan}, {Kapahi}, {Rao}, {Subrahmanya}, \& {Kulkarni}}]{Swarup_1991}
{Swarup}, G., {Ananthakrishnan}, S., {Kapahi}, V.~K., {et~al.} 1991, Current Science, 60, 95

\bibitem[{{Thompson} {et~al.}(2017){Thompson}, {Moran}, \& {Swenson}}]{Thompson_1986}
{Thompson}, A.~R., {Moran}, J.~M., \& {Swenson}, G.~W. 2017, {Interferometry and synthesis in radio astronomy} (Springer Cham)

\bibitem[{{Tingay} {et~al.}(2013){Tingay}, {Goeke}, {Bowman}, {Emrich}, {Ord}, {Mitchell}, {Morales}, {Booler}, {Crosse}, {Wayth}, {Lonsdale}, {Tremblay}, {Pallot}, {Colegate}, {Wicenec}, {Kudryavtseva}, {Arcus}, {Barnes}, {Bernardi}, {Briggs}, {Burns}, {Bunton}, {Cappallo}, {Corey}, {Deshpande}, {Desouza}, {Gaensler}, {Greenhill}, {Hall}, {Hazelton}, {Herne}, {Hewitt}, {Johnston-Hollitt}, {Kaplan}, {Kasper}, {Kincaid}, {Koenig}, {Kratzenberg}, {Lynch}, {Mckinley}, {Mcwhirter}, {Morgan}, {Oberoi}, {Pathikulangara}, {Prabu}, {Remillard}, {Rogers}, {Roshi}, {Salah}, {Sault}, {Udaya-Shankar}, {Schlagenhaufer}, {Srivani}, {Stevens}, {Subrahmanyan}, {Waterson}, {Webster}, {Whitney}, {Williams}, {Williams}, \& {Wyithe}}]{tingay13}
{Tingay}, S.~J., {Goeke}, R., {Bowman}, J.~D., {et~al.} 2013, \pasa, 30, e007, \dodoi{10.1017/pasa.2012.007}

\bibitem[{Trott {et~al.}(2019)}]{trott2019}
Trott, C.~M., {et~al.} 2019, Publ. Astron. Soc. Austral., 36, e023, \dodoi{10.1017/pasa.2019.15}

\bibitem[{Trott {et~al.}(2020)Trott, Jordan, Midgley, Barry, Greig, Pindor, Cook, Sleap, Tingay, Ung, Hancock, Williams, Bowman, Byrne, Chokshi, Hazelton, Hasegawa, Jacobs, Joseph, Li, Line, Lynch, McKinley, Mitchell, Morales, Ouchi, Pober, Rahimi, Takahashi, Wayth, Webster, Wilensky, Wyithe, Yoshiura, Zhang, \& Zheng}]{Trott2020}
Trott, C.~M., Jordan, C.~H., Midgley, S., {et~al.} 2020, \mnras, 493, 4711, \dodoi{10.1093/mnras/staa414}

\bibitem[{{van Haarlem, M. P.} {et~al.}(2013){van Haarlem, M. P.}, {Wise, M. W.}, {Gunst, A. W.}, {Heald, G.}, {McKean, J. P.}, {Hessels, J. W. T.}, {de Bruyn, A. G.}, {Nijboer, R.}, {Swinbank, J.}, {Fallows, R.}, {Brentjens, M.}, {Nelles, A.}, {Beck, R.}, {Falcke, H.}, {Fender, R.}, {H\"orandel, J.}, {Koopmans, L. V. E.}, {Mann, G.}, {Miley, G.}, {R\"ottgering, H.}, {Stappers, B. W.}, {Wijers, R. A. M. J.}, {Zaroubi, S.}, {van den Akker, M.}, {Alexov, A.}, {Anderson, J.}, {Anderson, K.}, {van Ardenne, A.}, {Arts, M.}, {Asgekar, A.}, {Avruch, I. M.}, {Batejat, F.}, {B\"ahren, L.}, {Bell, M. E.}, {Bell, M. R.}, {van Bemmel, I.}, {Bennema, P.}, {Bentum, M. J.}, {Bernardi, G.}, {Best, P.}, {B\^{\i}rzan, L.}, {Bonafede, A.}, {Boonstra, A.-J.}, {Braun, R.}, {Bregman, J.}, {Breitling, F.}, {van de Brink, R. H.}, {Broderick, J.}, {Broekema, P. C.}, {Brouw, W. N.}, {Br\"uggen, M.}, {Butcher, H. R.}, {van Cappellen, W.}, {Ciardi, B.}, {Coenen, T.}, {Conway, J.}, {Coolen, A.}, {Corstanje, A.}, {Damstra, S.}, {Davies,
  O.}, {Deller, A. T.}, {Dettmar, R.-J.}, {van Diepen, G.}, {Dijkstra, K.}, {Donker, P.}, {Doorduin, A.}, {Dromer, J.}, {Drost, M.}, {van Duin, A.}, {Eisl\"offel, J.}, {van Enst, J.}, {Ferrari, C.}, {Frieswijk, W.}, {Gankema, H.}, {Garrett, M. A.}, {de Gasperin, F.}, {Gerbers, M.}, {de Geus, E.}, {Grie\ss{}meier, J.-M.}, {Grit, T.}, {Gruppen, P.}, {Hamaker, J. P.}, {Hassall, T.}, {Hoeft, M.}, {Holties, H. A.}, {Horneffer, A.}, {van der Horst, A.}, {van Houwelingen, A.}, {Huijgen, A.}, {Iacobelli, M.}, {Intema, H.}, {Jackson, N.}, {Jelic, V.}, {de Jong, A.}, {Juette, E.}, {Kant, D.}, {Karastergiou, A.}, {Koers, A.}, {Kollen, H.}, {Kondratiev, V. I.}, {Kooistra, E.}, {Koopman, Y.}, {Koster, A.}, {Kuniyoshi, M.}, {Kramer, M.}, {Kuper, G.}, {Lambropoulos, P.}, {Law, C.}, {van Leeuwen, J.}, {Lemaitre, J.}, {Loose, M.}, {Maat, P.}, {Macario, G.}, {Markoff, S.}, {Masters, J.}, {McFadden, R. A.}, {McKay-Bukowski, D.}, {Meijering, H.}, {Meulman, H.}, {Mevius, M.}, {Middelberg, E.}, {Millenaar, R.}, {Miller-Jones, J.
  C. A.}, {Mohan, R. N.}, {Mol, J. D.}, {Morawietz, J.}, {Morganti, R.}, {Mulcahy, D. D.}, {Mulder, E.}, {Munk, H.}, {Nieuwenhuis, L.}, {van Nieuwpoort, R.}, {Noordam, J. E.}, {Norden, M.}, {Noutsos, A.}, {Offringa, A. R.}, {Olofsson, H.}, {Omar, A.}, {Orr\'u, E.}, {Overeem, R.}, {Paas, H.}, {Pandey-Pommier, M.}, {Pandey, V. N.}, {Pizzo, R.}, {Polatidis, A.}, {Rafferty, D.}, {Rawlings, S.}, {Reich, W.}, {de Reijer, J.-P.}, {Reitsma, J.}, {Renting, G. A.}, {Riemers, P.}, {Rol, E.}, {Romein, J. W.}, {Roosjen, J.}, {Ruiter, M.}, {Scaife, A.}, {van der Schaaf, K.}, {Scheers, B.}, {Schellart, P.}, {Schoenmakers, A.}, {Schoonderbeek, G.}, {Serylak, M.}, {Shulevski, A.}, {Sluman, J.}, {Smirnov, O.}, {Sobey, C.}, {Spreeuw, H.}, {Steinmetz, M.}, {Sterks, C. G. M.}, {Stiepel, H.-J.}, {Stuurwold, K.}, {Tagger, M.}, {Tang, Y.}, {Tasse, C.}, {Thomas, I.}, {Thoudam, S.}, {Toribio, M. C.}, {van der Tol, B.}, {Usov, O.}, {van Veelen, M.}, {van der Veen, A.-J.}, {ter Veen, S.}, {Verbiest, J. P. W.}, {Vermeulen, R.}, {Vermaas,
  N.}, {Vocks, C.}, {Vogt, C.}, {de Vos, M.}, {van der Wal, E.}, {van Weeren, R.}, {Weggemans, H.}, {Weltevrede, P.}, {White, S.}, {Wijnholds, S. J.}, {Wilhelmsson, T.}, {Wucknitz, O.}, {Yatawatta, S.}, {Zarka, P.}, {Zensus, A.}, \& {van Zwieten, J.}}]{haarlem}
{van Haarlem, M. P.}, {Wise, M. W.}, {Gunst, A. W.}, {et~al.} 2013, A\&A, 556, A2, \dodoi{10.1051/0004-6361/201220873}

\bibitem[{{Watkinson} {et~al.}(2021){Watkinson}, {Trott}, \& {Hothi}}]{Watkinson_2021}
{Watkinson}, C.~A., {Trott}, C.~M., \& {Hothi}, I. 2021, \mnras, 501, 367, \dodoi{10.1093/mnras/staa3677}

\bibitem[{Yoshiura {et~al.}(2026)Yoshiura, Ito, Null, Takahashi, Trott, Hidayat, \& Pindor}]{yoshiora2026}
Yoshiura, S., Ito, T., Null, D., {et~al.} 2026, Publications of the Astronomical Society of Japan, psag110, \dodoi{10.1093/pasj/psag110}

\bibitem[{{Yoshiura} {et~al.}(2015){Yoshiura}, {Shimabukuro}, {Takahashi}, {Momose}, {Nakanishi}, \& {Imai}}]{Yoshiura_2015}
{Yoshiura}, S., {Shimabukuro}, H., {Takahashi}, K., {et~al.} 2015, \mnras, 451, 266, \dodoi{10.1093/mnras/stv855}

\end{thebibliography}
\bibliographystyle{aasjournal}

\appendix

\section{Estimates of 3D spherical Bispectrum for different triangle shapes and sizes}
\label{appendix:3dbs_kmut}

\begin{figure*}[htbp]
    \centering
    \includegraphics[width=\textwidth]{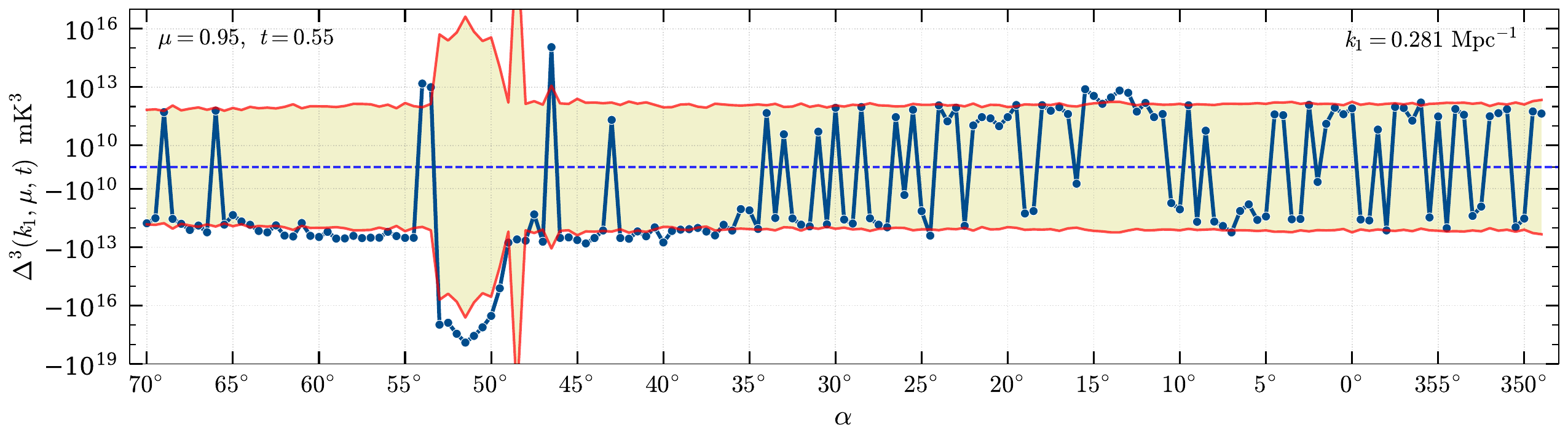}
    \caption{Variation of $\Delta^{3}(k_1,\mu,t)$ with right ascension $\alpha$, for  $k_1 = 0.281\,\mathrm{Mpc}^{-1}$.    The solid blue curve is the estimated $\Delta^{3}$. The yellow shaded region with red boundaries denotes $0 \pm 2\sigma$, the expected statistical fluctuations due to system noise. In this case we have estimates for a single shape with $(\mu,t)=(0.95,0.55)$ as indicated in the figure. Note the severe foreground contamination around $\alpha \approx 50.5^{\circ}$ due to the bright radio source Fornax~A appearing within the main lobe of the MWA primary beam.}
    \label{fig:3dbs_k1}
\end{figure*}

\begin{figure*}[htbp]
    \centering
    \includegraphics[width=\textwidth]{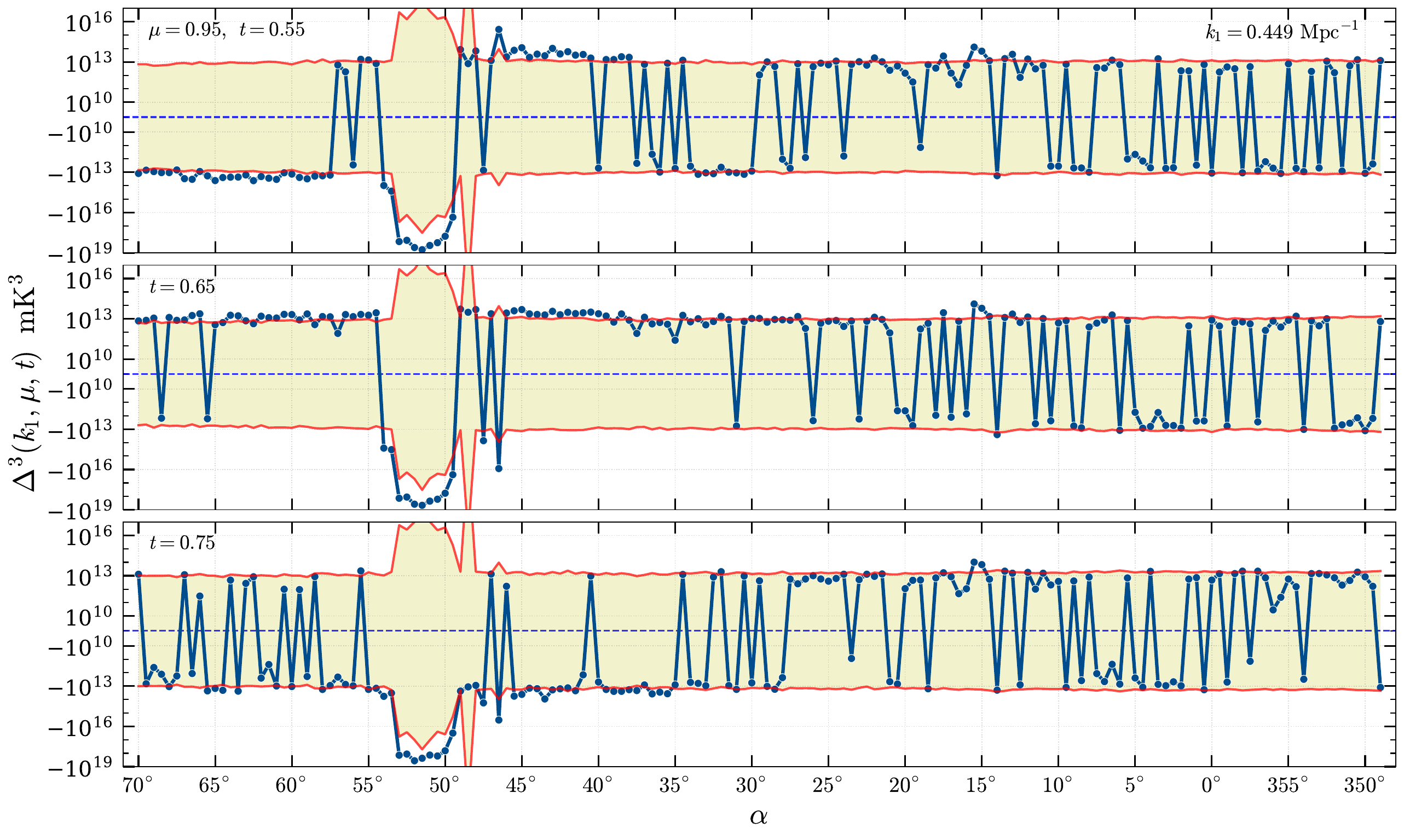}
    \caption{Same as Fig.~\ref{fig:3dbs_k1}, but for $k_1=0.449\,\mathrm{Mpc}^{-1}$. Each panel corresponds to a different shape as indicated in the figure.}
    \label{fig:3dbs_k2}
\end{figure*}

\begin{figure*}[htbp]
    \centering
    \includegraphics[width=\textwidth]{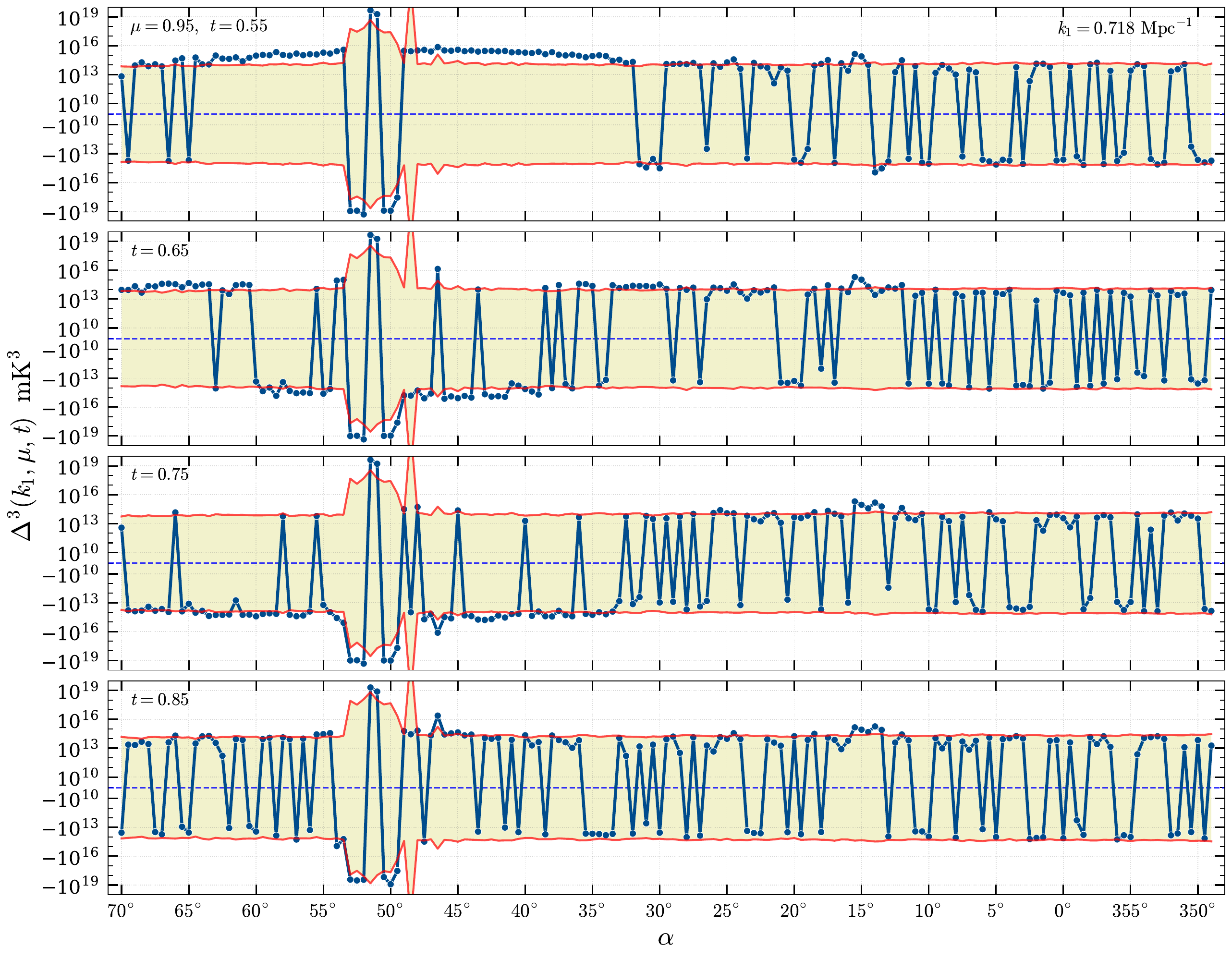}
    \caption{Same as Fig.~\ref{fig:3dbs_k1}, but for $k_1=0.718\,\mathrm{Mpc}^{-1}$. Each panel corresponds to a different shape as indicated in the figure.}
    \label{fig:3dbs_k3}
\end{figure*}

\begin{figure*}[htbp]
    \centering
    \includegraphics[width=\textwidth]{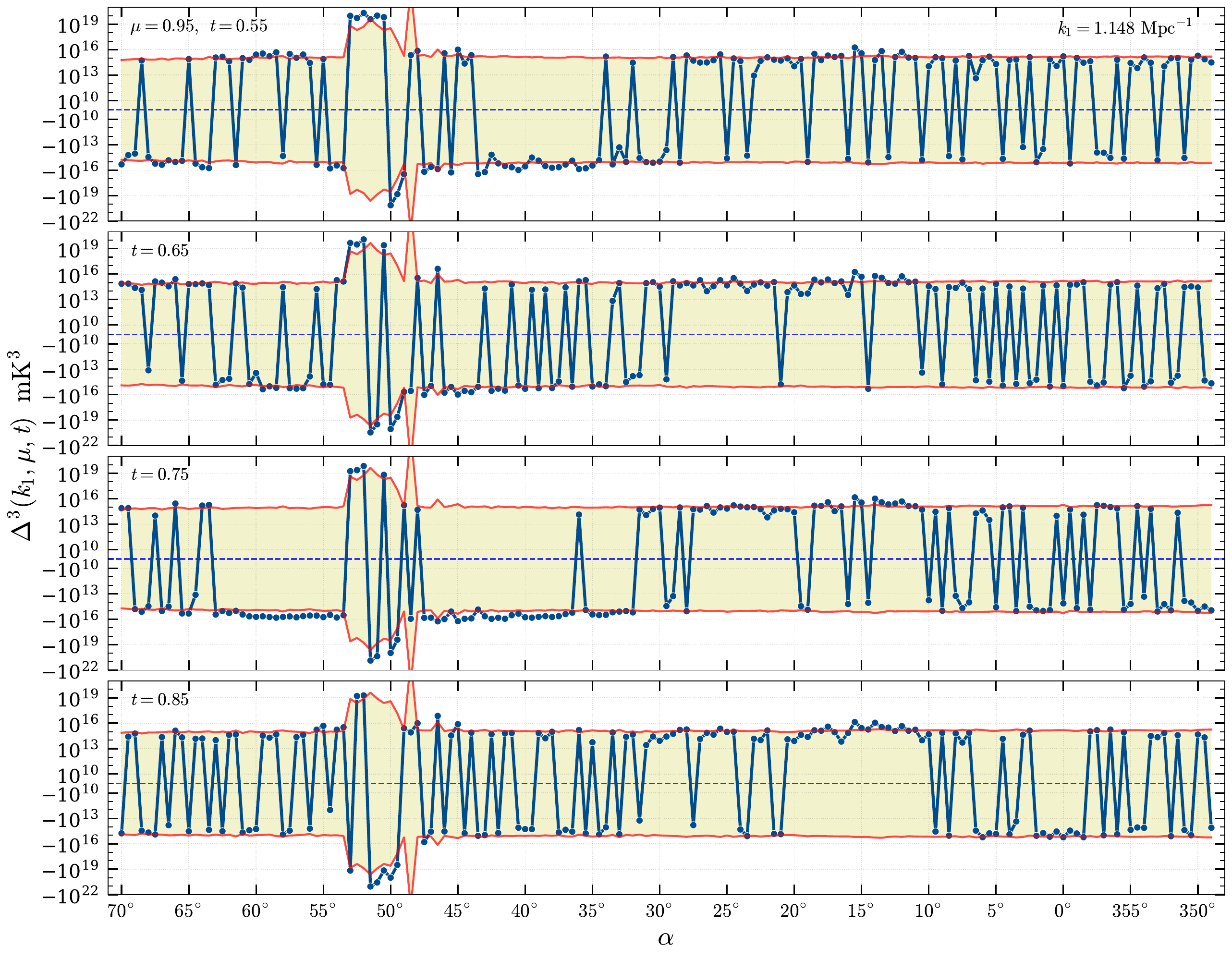}
    \caption{Same as Fig.~\ref{fig:3dbs_k1}, but for $k_1=1.148\,\mathrm{Mpc}^{-1}$. Each panel corresponds to a different shape as indicated in the figure.}
    \label{fig:3dbs_k4}
\end{figure*}

\begin{figure*}[htbp]
    \centering
    \includegraphics[width=\textwidth]{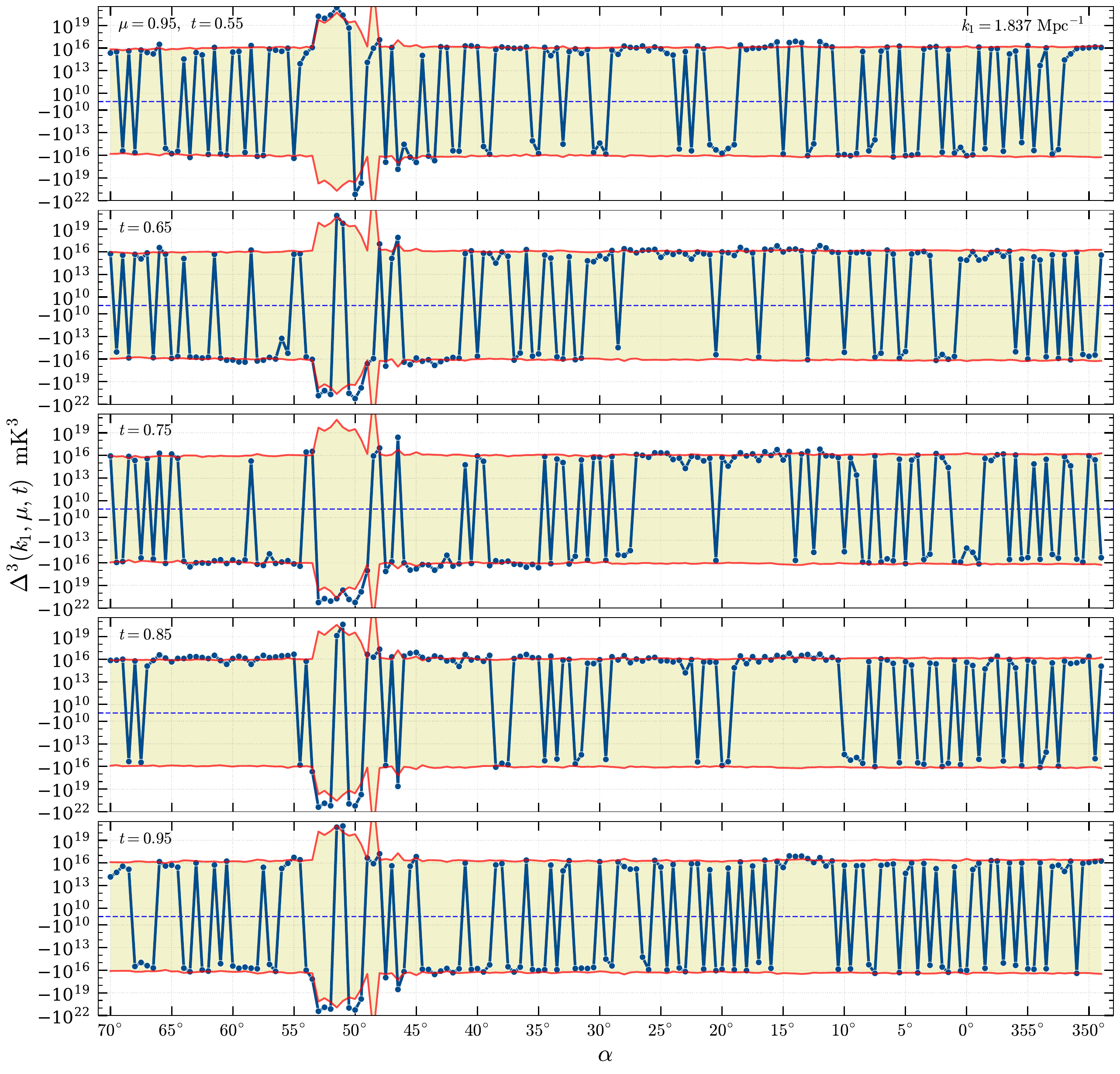}
    \caption{Same as Fig.~\ref{fig:3dbs_k1}, but for $k_1=1.837 \,\mathrm{Mpc}^{-1}$. Each panel corresponds to a different shape as indicated in the figure.}
    \label{fig:3dbs_k5}
\end{figure*}

\begin{figure*}[htbp]
    \centering
    \includegraphics[width=\textwidth]{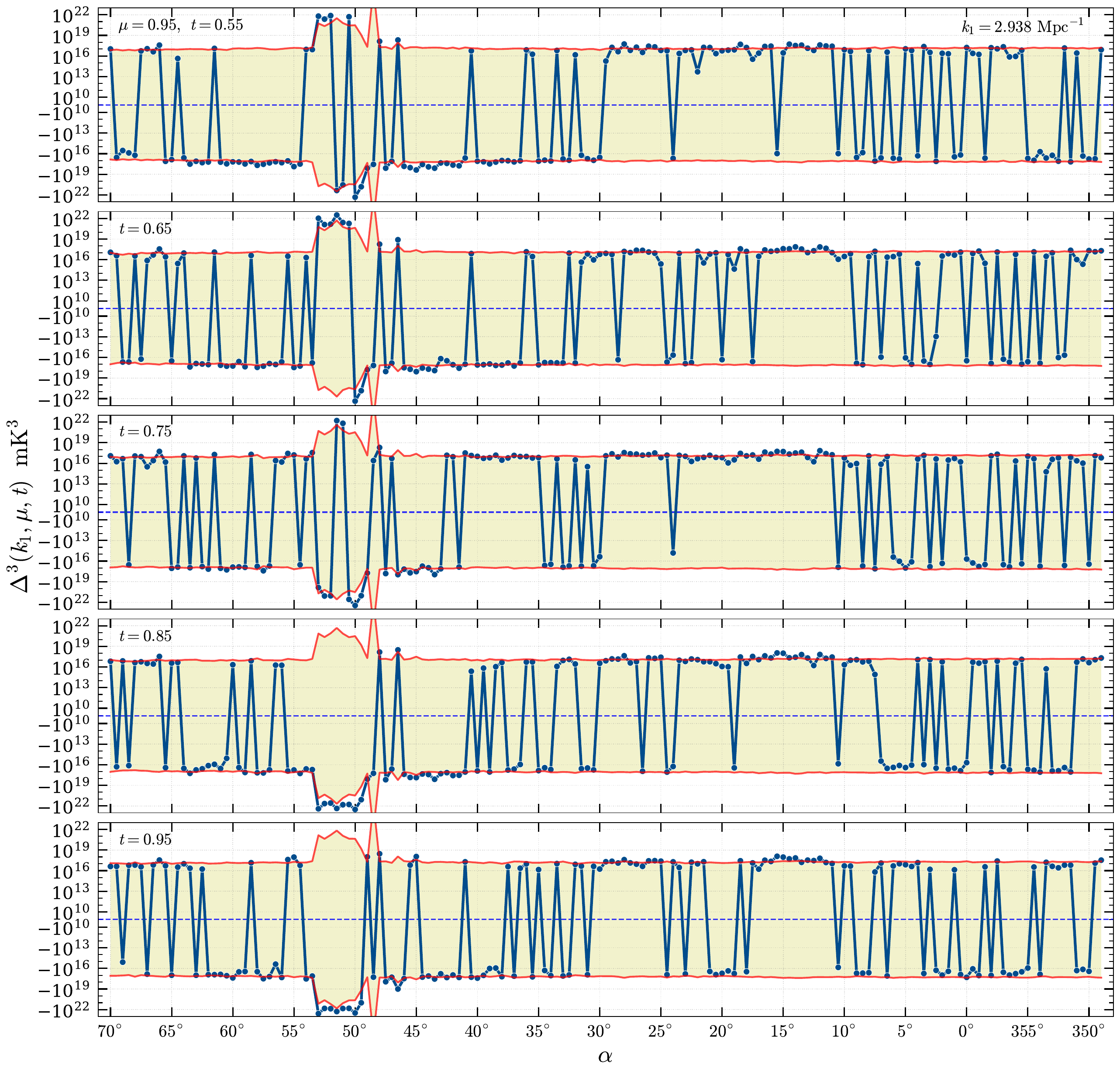}
    \caption{Same as Fig.~\ref{fig:3dbs_k1}, but for $k_1=2.938 \,\mathrm{Mpc}^{-1}$. Each panel corresponds to a different shape as indicated in the figure.}
    \label{fig:3dbs_k6}
\end{figure*}

\begin{figure*}[htbp]
    \centering
    \includegraphics[width=\textwidth]{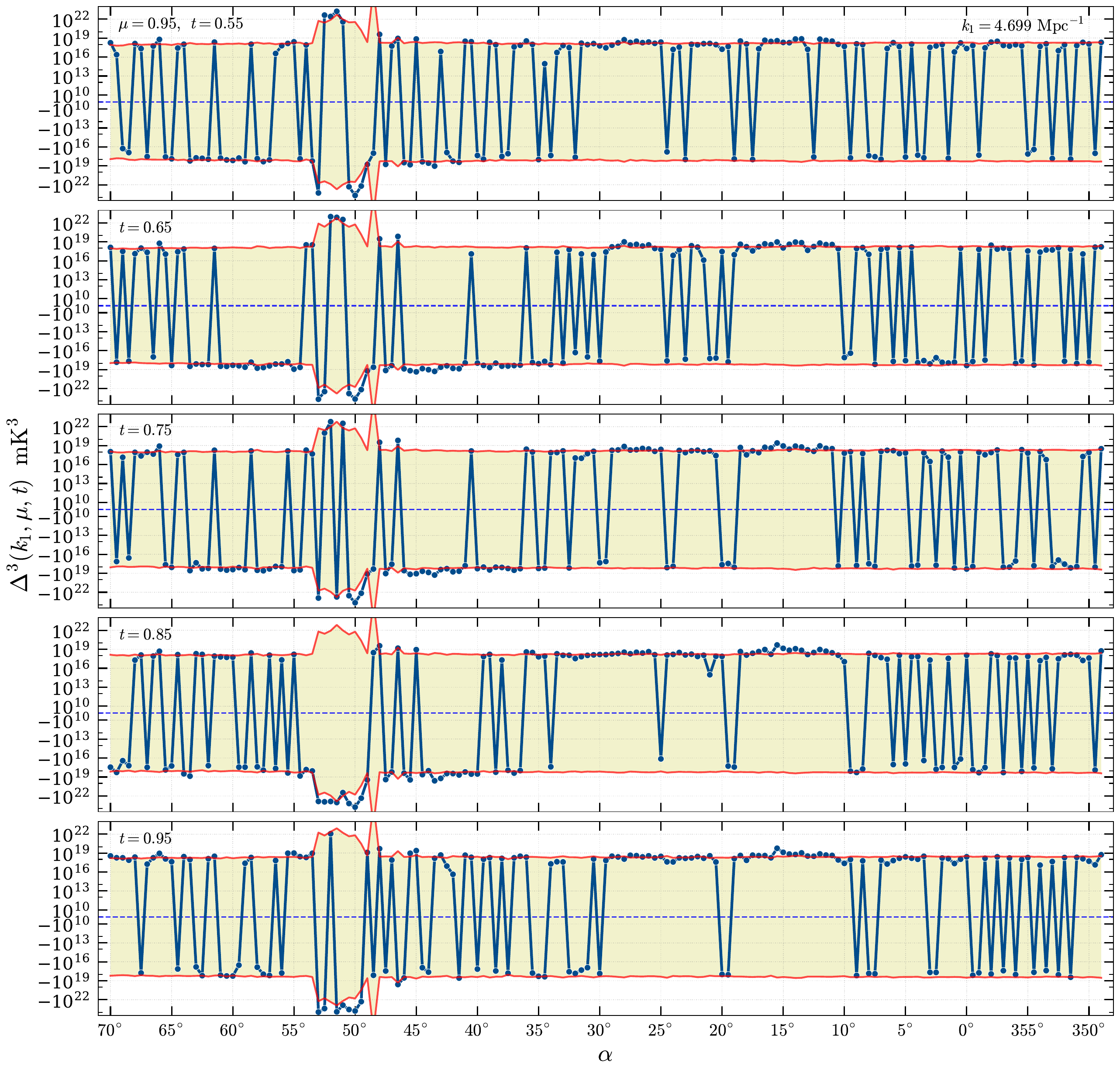}
    \caption{Same as Fig.~\ref{fig:3dbs_k1}, but for $k_1=4.699\,\mathrm{Mpc}^{-1}$. Each panel corresponds to a different shape as indicated in the figure.}
    \label{fig:3dbs_k7}
\end{figure*}

\vspace{0.5cm}
Figures~\ref{fig:3dbs_k1} to  \ref{fig:3dbs_k7} all show the variation of $\Delta^3(k_1,\mu,t)$ with $\alpha$. Each figure corresponds to a different $k$ bin, for which each panel corresponds to a different triangle shape. Note that we 
obtain estimates of  $\Delta^3(k_1,\mu,t)$ for only a limited number of triangle shapes, all of which correspond to the bin centered at $\mu=0.95$. The various panels here correspond to different values of $t$. 

\vspace{3cm}

\clearpage

\section{Pointing centers used for incoherent combination}
\label{appendix:table_PCs}
Table~\ref{tab:pcs_used} lists the RAs of the PCs that were used for the incoherent combination for each $k$ bin. In all cases, except for $k_1=1.837~{\rm Mpc}^{-1}$, we have used all the PCs that satisfy the condition $|\Delta^3|\le 2 \sigma$. For $k_1=1.837~{\rm Mpc}^{-1}$, we sorted the PCs that satisfy $|\Delta^3|\le 2 \sigma$ in order of increasing $\sigma$,  and excluded the last $26$ PCs from the incoherent combination. 

\begin{table*}[htbp]
\centering
\caption{Right ascensions ($\alpha$, in degrees) of the PCs used for the incoherent combination at each $k$ bin. In each row, the PCs are listed in order of increasing $\sigma$.}
\label{tab:pcs_used}
\renewcommand{\arraystretch}{1.2}
\small
\begin{tabular}{|c|c|}
\hline
$k_1\,(\mathrm{Mpc}^{-1})$ & RA of the PCs \\
\hline
0.281 & \parbox[t]{0.78\linewidth}{\raggedright $69^{\circ}$, $64.5^{\circ}$, $68^{\circ}$, $70^{\circ}$, $69.5^{\circ}$, $61^{\circ}$, $65^{\circ}$, $66^{\circ}$, $67^{\circ}$, $30.5^{\circ}$, $64^{\circ}$, $21.5^{\circ}$, $20^{\circ}$, $33^{\circ}$, $31.5^{\circ}$, $16^{\circ}$, $32.5^{\circ}$, $29.5^{\circ}$, $26.5^{\circ}$, $22.5^{\circ}$, $26^{\circ}$, $40.5^{\circ}$, $352^{\circ}$, $35.5^{\circ}$, $16.5^{\circ}$, $27^{\circ}$, $23.5^{\circ}$, $68.5^{\circ}$, $28.5^{\circ}$, $30^{\circ}$, $35^{\circ}$, $31^{\circ}$, $27.5^{\circ}$, $34^{\circ}$, $12.5^{\circ}$, $0.5^{\circ}$, $20.5^{\circ}$, $8^{\circ}$, $17.5^{\circ}$, $23^{\circ}$, $18.5^{\circ}$, $21^{\circ}$, $25^{\circ}$, $11^{\circ}$, $359.5^{\circ}$, $36.5^{\circ}$, $358.5^{\circ}$, $29^{\circ}$, $34.5^{\circ}$, $10.5^{\circ}$, $8.5^{\circ}$, $18^{\circ}$, $350^{\circ}$, $351^{\circ}$, $19^{\circ}$, $357^{\circ}$, $9^{\circ}$, $2^{\circ}$, $11.5^{\circ}$, $1^{\circ}$, $24^{\circ}$, $6.5^{\circ}$, $25.5^{\circ}$, $355.5^{\circ}$, $10^{\circ}$, $7.5^{\circ}$, $33.5^{\circ}$, $3^{\circ}$, $6^{\circ}$, $5^{\circ}$, $358^{\circ}$, $353^{\circ}$, $1.5^{\circ}$, $22^{\circ}$, $351.5^{\circ}$, $32^{\circ}$, $355^{\circ}$, $359^{\circ}$, $43^{\circ}$, $28^{\circ}$, $17^{\circ}$, $357.5^{\circ}$, $5.5^{\circ}$, $356.5^{\circ}$, $354^{\circ}$, $352.5^{\circ}$, $2.5^{\circ}$, $354.5^{\circ}$, $353.5^{\circ}$, $14.5^{\circ}$, $4^{\circ}$, $350.5^{\circ}$, $4.5^{\circ}$, $47.5^{\circ}$, $3.5^{\circ}$, $0^{\circ}$, $349.5^{\circ}$, $349^{\circ}$, $48.5^{\circ}$} \\
\hline
0.449 & \parbox[t]{0.78\linewidth}{\raggedright $68^{\circ}$, $69.5^{\circ}$, $70^{\circ}$, $63^{\circ}$, $67.5^{\circ}$, $35^{\circ}$, $69^{\circ}$, $36^{\circ}$, $62^{\circ}$, $60^{\circ}$, $34^{\circ}$, $30^{\circ}$, $66^{\circ}$, $60.5^{\circ}$, $64^{\circ}$, $37.5^{\circ}$, $28.5^{\circ}$, $25.5^{\circ}$, $20^{\circ}$, $33.5^{\circ}$, $63.5^{\circ}$, $31.5^{\circ}$, $26^{\circ}$, $22.5^{\circ}$, $59^{\circ}$, $19.5^{\circ}$, $56^{\circ}$, $57.5^{\circ}$, $19^{\circ}$, $27.5^{\circ}$, $29.5^{\circ}$, $21^{\circ}$, $20.5^{\circ}$, $29^{\circ}$, $32^{\circ}$, $23^{\circ}$, $35.5^{\circ}$, $57^{\circ}$, $18.5^{\circ}$, $17^{\circ}$, $356.5^{\circ}$, $8^{\circ}$, $357.5^{\circ}$, $16^{\circ}$, $23.5^{\circ}$, $36.5^{\circ}$, $11.5^{\circ}$, $18^{\circ}$, $10^{\circ}$, $24^{\circ}$, $25^{\circ}$, $30.5^{\circ}$, $16.5^{\circ}$, $10.5^{\circ}$, $26.5^{\circ}$, $359^{\circ}$, $9.5^{\circ}$, $3^{\circ}$, $32.5^{\circ}$, $359.5^{\circ}$, $353.5^{\circ}$, $358.5^{\circ}$, $9^{\circ}$, $358^{\circ}$, $356^{\circ}$, $12.5^{\circ}$, $5^{\circ}$, $1^{\circ}$, $58^{\circ}$, $357^{\circ}$, $3.5^{\circ}$, $2.5^{\circ}$, $7.5^{\circ}$, $349.5^{\circ}$, $11^{\circ}$, $33^{\circ}$, $0.5^{\circ}$, $351.5^{\circ}$, $355.5^{\circ}$, $4.5^{\circ}$, $354.5^{\circ}$, $28^{\circ}$, $5.5^{\circ}$, $1.5^{\circ}$, $352^{\circ}$, $4^{\circ}$, $354^{\circ}$, $2^{\circ}$, $355^{\circ}$, $351^{\circ}$, $8.5^{\circ}$, $0^{\circ}$, $353^{\circ}$, $350^{\circ}$, $7^{\circ}$, $350.5^{\circ}$, $6^{\circ}$, $349^{\circ}$, $48.5^{\circ}$} \\
\hline
0.718 & \parbox[t]{0.78\linewidth}{\raggedright $69.5^{\circ}$, $70^{\circ}$, $31^{\circ}$, $30^{\circ}$, $21^{\circ}$, $18^{\circ}$, $27^{\circ}$, $20.5^{\circ}$, $357^{\circ}$, $11^{\circ}$, $16^{\circ}$, $23.5^{\circ}$, $26.5^{\circ}$, $20^{\circ}$, $31.5^{\circ}$, $17^{\circ}$, $22.5^{\circ}$, $1^{\circ}$, $19.5^{\circ}$, $24^{\circ}$, $349.5^{\circ}$, $5.5^{\circ}$, $8^{\circ}$, $19^{\circ}$, $9^{\circ}$, $5^{\circ}$, $10.5^{\circ}$, $29^{\circ}$, $7^{\circ}$, $354.5^{\circ}$, $355^{\circ}$, $11.5^{\circ}$, $356.5^{\circ}$, $9.5^{\circ}$, $359.5^{\circ}$, $12.5^{\circ}$, $13^{\circ}$, $2.5^{\circ}$, $352.5^{\circ}$, $8.5^{\circ}$, $10^{\circ}$, $6.5^{\circ}$, $350.5^{\circ}$, $354^{\circ}$, $359^{\circ}$, $0.5^{\circ}$, $4.5^{\circ}$, $1.5^{\circ}$, $356^{\circ}$, $358^{\circ}$, $358.5^{\circ}$, $2^{\circ}$, $355.5^{\circ}$, $3^{\circ}$, $351.5^{\circ}$, $4^{\circ}$, $6^{\circ}$, $28^{\circ}$, $350^{\circ}$, $7.5^{\circ}$, $349^{\circ}$, $353.5^{\circ}$, $3.5^{\circ}$, $353^{\circ}$, $0^{\circ}$, $48.5^{\circ}$} \\
\hline
1.148 & \parbox[t]{0.78\linewidth}{\raggedright $68^{\circ}$, $70^{\circ}$, $68.5^{\circ}$, $69^{\circ}$, $66.5^{\circ}$, $32^{\circ}$, $64.5^{\circ}$, $62.5^{\circ}$, $62^{\circ}$, $67.5^{\circ}$, $65^{\circ}$, $31.5^{\circ}$, $30.5^{\circ}$, $19.5^{\circ}$, $29.5^{\circ}$, $28.5^{\circ}$, $30^{\circ}$, $21^{\circ}$, $16^{\circ}$, $20.5^{\circ}$, $21.5^{\circ}$, $31^{\circ}$, $356^{\circ}$, $25^{\circ}$, $23^{\circ}$, $55^{\circ}$, $359.5^{\circ}$, $27.5^{\circ}$, $22.5^{\circ}$, $5.5^{\circ}$, $26^{\circ}$, $8^{\circ}$, $10.5^{\circ}$, $20^{\circ}$, $358^{\circ}$, $19^{\circ}$, $23.5^{\circ}$, $9^{\circ}$, $353.5^{\circ}$, $352^{\circ}$, $351^{\circ}$, $26.5^{\circ}$, $8.5^{\circ}$, $2.5^{\circ}$, $4^{\circ}$, $10^{\circ}$, $355^{\circ}$, $0.5^{\circ}$, $357^{\circ}$, $6.5^{\circ}$, $1.5^{\circ}$, $357.5^{\circ}$, $354^{\circ}$, $351.5^{\circ}$, $4.5^{\circ}$, $350.5^{\circ}$, $359^{\circ}$, $33^{\circ}$, $352.5^{\circ}$, $9.5^{\circ}$, $5^{\circ}$, $3.5^{\circ}$, $350^{\circ}$, $354.5^{\circ}$, $6^{\circ}$, $358.5^{\circ}$, $3^{\circ}$, $14.5^{\circ}$, $7^{\circ}$, $0^{\circ}$, $349^{\circ}$, $7.5^{\circ}$, $353^{\circ}$, $349.5^{\circ}$, $48.5^{\circ}$} \\
\hline
1.837 & \parbox[t]{0.78\linewidth}{\raggedright $65.5^{\circ}$, $67.5^{\circ}$, $69^{\circ}$, $64.5^{\circ}$, $68.5^{\circ}$, $62.5^{\circ}$, $37.5^{\circ}$, $69.5^{\circ}$, $67^{\circ}$, $60^{\circ}$, $59.5^{\circ}$, $65^{\circ}$, $68^{\circ}$, $39^{\circ}$, $61^{\circ}$, $66.5^{\circ}$, $31.5^{\circ}$, $63^{\circ}$, $56.5^{\circ}$, $21^{\circ}$, $35.5^{\circ}$, $64^{\circ}$, $34^{\circ}$, $30.5^{\circ}$, $57.5^{\circ}$, $31^{\circ}$, $19.5^{\circ}$, $21.5^{\circ}$, $62^{\circ}$, $358.5^{\circ}$, $23^{\circ}$, $3^{\circ}$, $10.5^{\circ}$, $36.5^{\circ}$, $20^{\circ}$, $20.5^{\circ}$, $32.5^{\circ}$, $29.5^{\circ}$, $38^{\circ}$, $34.5^{\circ}$, $17^{\circ}$, $9.5^{\circ}$, $9^{\circ}$, $24^{\circ}$, $7^{\circ}$, $19^{\circ}$, $28.5^{\circ}$, $23.5^{\circ}$, $352.5^{\circ}$, $2^{\circ}$, $22.5^{\circ}$, $0.5^{\circ}$, $355.5^{\circ}$, $32^{\circ}$, $37^{\circ}$, $352^{\circ}$, $54.5^{\circ}$, $5^{\circ}$, $354^{\circ}$, $38.5^{\circ}$, $356^{\circ}$, $1.5^{\circ}$, $10^{\circ}$, $33.5^{\circ}$, $4^{\circ}$, $8.5^{\circ}$, $42.5^{\circ}$, $359^{\circ}$, $351.5^{\circ}$} \\
\hline
2.938 & \parbox[t]{0.78\linewidth}{\raggedright $68.5^{\circ}$, $69^{\circ}$, $67^{\circ}$, $69.5^{\circ}$, $64.5^{\circ}$, $65.5^{\circ}$, $31.5^{\circ}$, $38.5^{\circ}$, $33.5^{\circ}$, $32.5^{\circ}$, $20^{\circ}$, $24^{\circ}$, $33^{\circ}$, $54^{\circ}$, $37.5^{\circ}$, $35.5^{\circ}$, $9^{\circ}$, $19.5^{\circ}$, $10.5^{\circ}$, $356.5^{\circ}$, $358.5^{\circ}$, $30^{\circ}$, $359.5^{\circ}$, $31^{\circ}$, $23^{\circ}$, $36.5^{\circ}$, $19^{\circ}$, $24.5^{\circ}$, $8.5^{\circ}$, $354.5^{\circ}$, $356^{\circ}$, $40.5^{\circ}$, $0.5^{\circ}$, $4^{\circ}$, $4.5^{\circ}$, $32^{\circ}$, $3^{\circ}$, $17^{\circ}$, $41^{\circ}$, $1.5^{\circ}$, $359^{\circ}$, $2.5^{\circ}$, $1^{\circ}$, $7^{\circ}$, $352.5^{\circ}$, $355^{\circ}$, $357^{\circ}$, $5^{\circ}$, $351^{\circ}$, $6^{\circ}$, $351.5^{\circ}$, $355.5^{\circ}$, $5.5^{\circ}$, $6.5^{\circ}$, $2^{\circ}$, $352^{\circ}$, $350^{\circ}$, $3.5^{\circ}$, $7.5^{\circ}$, $350.5^{\circ}$, $353.5^{\circ}$, $358^{\circ}$, $0^{\circ}$, $353^{\circ}$, $52^{\circ}$, $48.5^{\circ}$} \\
\hline
4.699 & \parbox[t]{0.78\linewidth}{\raggedright $69^{\circ}$, $68.5^{\circ}$, $65.5^{\circ}$, $32^{\circ}$, $56^{\circ}$, $64^{\circ}$, $24^{\circ}$, $20^{\circ}$, $34^{\circ}$, $20.5^{\circ}$, $63^{\circ}$, $32.5^{\circ}$, $67^{\circ}$, $29.5^{\circ}$, $19.5^{\circ}$, $30^{\circ}$, $34.5^{\circ}$, $10.5^{\circ}$, $23^{\circ}$, $356.5^{\circ}$, $40.5^{\circ}$, $9^{\circ}$, $8^{\circ}$, $3.5^{\circ}$, $4^{\circ}$, $5^{\circ}$, $10^{\circ}$, $354.5^{\circ}$, $351.5^{\circ}$, $349.5^{\circ}$, $358.5^{\circ}$, $39^{\circ}$, $357^{\circ}$, $36.5^{\circ}$, $2.5^{\circ}$, $4.5^{\circ}$, $19^{\circ}$, $356^{\circ}$, $24.5^{\circ}$, $41^{\circ}$, $351^{\circ}$, $55^{\circ}$, $6.5^{\circ}$, $7^{\circ}$, $33^{\circ}$, $354^{\circ}$, $353.5^{\circ}$, $9.5^{\circ}$, $1.5^{\circ}$, $1^{\circ}$, $359^{\circ}$, $353^{\circ}$, $2^{\circ}$, $350^{\circ}$, $359.5^{\circ}$, $3^{\circ}$, $352.5^{\circ}$, $350.5^{\circ}$, $355^{\circ}$, $352^{\circ}$, $6^{\circ}$, $355.5^{\circ}$, $0^{\circ}$, $7.5^{\circ}$, $48.5^{\circ}$} \\
\hline
\end{tabular}
\end{table*}

\label{lastpage}
\end{document}